\documentclass[sigconf]{acmart}
\usepackage{multirow}
\usepackage{nicematrix}
\usepackage{caption}
\usepackage{subcaption}
\usepackage{tabularx}
\usepackage{longtable}
\usepackage{ltcaption}
\usepackage{colortbl}

\usepackage{stfloats}

\AtBeginDocument{%
  \providecommand\BibTeX{{%
    \normalfont B\kern-0.5em{\scshape i\kern-0.25em b}\kern-0.8em\TeX}}}

\usepackage{soul}

\copyrightyear{2024}
\acmYear{2024}
\setcopyright{rightsretained}
\acmConference[ASSETS '24]{The 26th International ACM SIGACCESS Conference on Computers and Accessibility}{October 27--30, 2024}{St. John's, NL, Canada}
\acmBooktitle{The 26th International ACM SIGACCESS Conference on Computers and Accessibility (ASSETS '24), October 27--30, 2024, St. John's, NL, Canada}
\acmDOI{10.1145/3663548.3675638}
\acmISBN{979-8-4007-0677-6/24/10}

\begin{document}
\newcommand{\camera}[1]{\textcolor{blue}{[Camera Ready: #1]}}

\title{WheelPoser: Sparse-IMU Based Body Pose Estimation for~Wheelchair Users}




\author{Yunzhi Li}
\affiliation{%
    \institution{Carnegie Mellon University}
    \city{Pittsburgh}
    \state{PA}
    \country{USA}
 }
 \email{yunzhil@cs.cmu.edu}

\author{Vimal Mollyn}
\affiliation{
  \institution{Carnegie Mellon University}
  \city{Pittsburgh}
  \state{PA}
  \country{USA}
}
\email{vmollyn@cs.cmu.edu}

\author{Kuang Yuan}
\affiliation{
  \institution{Carnegie Mellon University}
  \city{Pittsburgh}
  \state{PA}
  \country{USA}
}
\email{kuangy@andrew.cmu.edu }

\author{Patrick Carrington}
\affiliation{%
    \institution{Carnegie Mellon University}
    \city{Pittsburgh}
    \state{PA}
    \country{USA}
 }
 \email{pcarrington@cmu.edu}

\renewcommand{\shortauthors}{Li et al.}

\begin{abstract}
Despite researchers having extensively studied various ways to track body pose on-the-go, most prior work does not take into account wheelchair users, leading to poor tracking performance. Wheelchair users could greatly benefit from this pose information to prevent injuries, monitor their health, identify environmental accessibility barriers, and interact with gaming and VR experiences. In this work, we present WheelPoser, a real-time pose estimation system specifically designed for wheelchair users. Our system uses only four strategically placed IMUs on the user's body and wheelchair, making it far more practical than prior systems using cameras and dense IMU arrays. WheelPoser is able to track a wheelchair user's pose with a mean joint angle error of $14.30^\circ$ and a mean joint position error of $6.74$~cm, more than three times better than similar systems using sparse IMUs. To train our system, we collect a novel WheelPoser-IMU dataset, consisting of 167 minutes of paired IMU sensor and motion capture data of people in wheelchairs, including wheelchair-specific motions such as propulsion and pressure relief. 
Finally, we explore the potential application space enabled by our system and discuss future opportunities. Open-source code, models, and dataset can be found here: \url{https://github.com/axle-lab/WheelPoser}.
\end{abstract}





\begin{teaserfigure}
    \includegraphics[width=\textwidth]{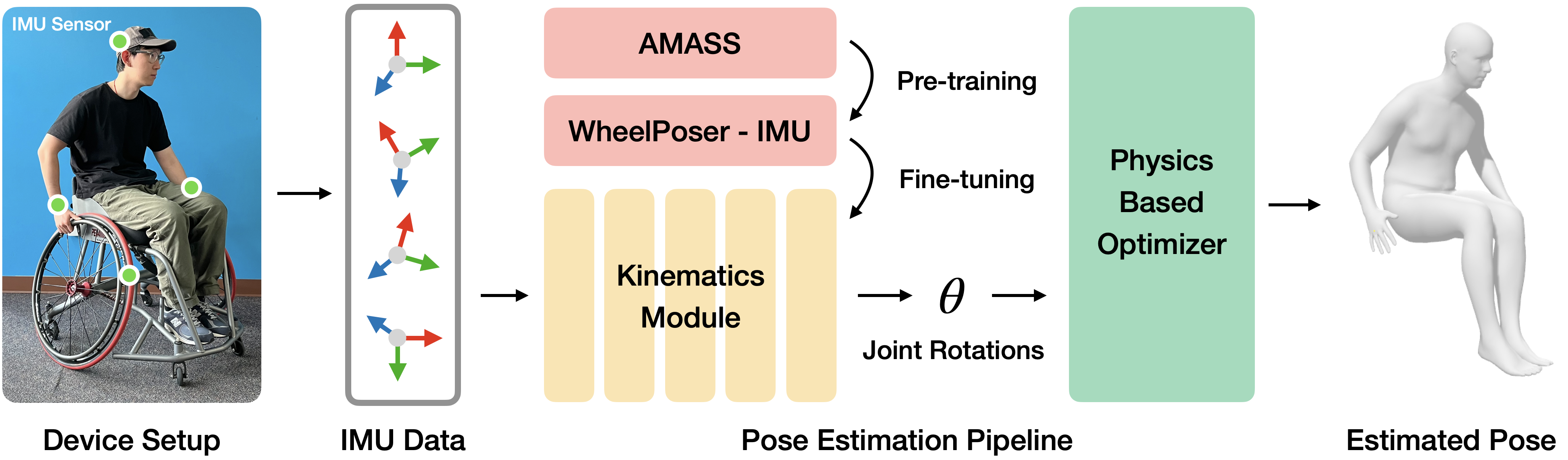}
    \vspace{1pt}
    \caption{WheelPoser uses a sparse-IMU setup for practical and effective pose estimation of wheelchair users. 
    It comprises a learning-based kinematics module that maps sparse IMU measurements to joint rotations and a physics-based optimizer for motion refinement. 
    Our models are pre-trained on the AMASS dataset and fine-tuned on our WheelPoser-IMU dataset leading to state-of-the-art pose estimation performance for wheelchair users.}
    \Description{This figure is a flow diagram comprising five major components that describe the pose estimation pipeline of WheelPoser. Starting from the left, there is an individual seated in a manual wheelchair with four IMU sensors attached to both their body and the wheelchair. Next, a vertical series of colored arrows represents the data from each IMU sensor. Following this, a yellow block labeled “Kinematics Module” takes the IMU data as input and outputs estimated joint rotations. Above this, two red blocks represent the datasets used to train and fine-tune this kinematics module, namely AMASS and WheelPoser-IMU, respectively. Further to the right, a green-colored block indicates the final stage where the pose data is refined by a physics-based optimizer. Finally, on the far right, a human mesh in a sitting posture represents the pose estimated by WheelPoser.}
    \label{fig:pipeline}
    \end{teaserfigure}
\maketitle

\section{Introduction}
The upper body pose of people who use wheelchairs contains rich and valuable information about their behavior, activities, and interactions with the environment. Being able to track and understand these poses on a daily basis holds immense potential to enhance various aspects of wheelchair users' lives and open up new applications and experiences. For instance, tracking the day-to-day upper body mechanics of a wheelchair user can be used to promote their health awareness and facilitate injury prevention~\cite{li2023breaking}. Long-term monitoring of wheelchair users' daily movement patterns across various contexts can also be instrumental in wheelchair fitting and guiding novel design for maximized comfort and mobility~\cite{kon2015development}, as well as identifying environmental barriers for accessible modifications~\cite{jarosz1996determination}. More broadly, pose tracking allows for precise analytics and training for adaptive sports, while also affording new interaction modalities for consumer-level applications such as inclusive gaming and VR experiences~\cite{gerling2020virtual, mott2020just}.

Despite this promising potential, there is still no system that achieves practical upper body pose estimation for wheelchair users across diverse contexts.
Most commonly, wheelchair users' poses are captured using cameras, such as optical cameras in marker-based motion capture systems (e.g., Vicon~\cite{ViconAwa72:online}) and RGB or RGB-D cameras~\cite{cotton2020kinematic,wei2016can,hwang2017feasibility, milgrom2016reliability, rammer2018assessment, huang2024wheelpose}. Yet, these systems are privacy-invasive, sensitive to occlusion, and most importantly do not work on the go.
In contrast, pose estimation using inertial measurement units (IMUs) is portable, able to work in any environment, insusceptible to visual occlusions, and more privacy-preserving, making it a more suitable solution for everyday consumer-level usage. However, existing IMU-based motion capture systems require instrumenting the wheelchair user with dense IMU arrays on each joint, totaling 11 or more sensors, to effectively estimate upper body poses~\cite{starrs2012biomechanical, kondo2024kinematic}, significantly reducing their practicality and usability.

In this work, we present WheelPoser, a motion capture system that achieves real-time upper body pose estimation for wheelchair users using a total of 4 strategically positioned IMUs on both the wheelchair and the user's body. Drawing inspiration from previous studies that use sparse IMUs for pose estimation of people without motor impairments~\cite{huang2018deep,yi2021transpose,jiang2022transformer,mollyn2023imuposer}, WheelPoser employs a learning-based kinematics module to map sparse inertial measurements to a broader set of upper body joint angles. 
Unfortunately, there is no publicly available motion capture dataset of wheelchair users, let alone one on a large scale, that would allow us to directly train such a model. 
To address this, we propose a data synthesis and pose estimation pipeline tailored for wheelchair users. Specifically, we first train a base model by leveraging the diverse upper body motions in existing datasets collected from people without motor impairments~\cite{mahmood2019amass}. 
We then fine-tune this pre-trained model on our newly collected WheelPoser-IMU dataset, comprising 167 minutes of typical on-wheelchair motions with IMU sensor data from 14 participants, in order to bridge the gap between existing motions and on-wheelchair motions.
Additionally, we propose a physics-based optimization module tailored for on-wheelchair motions, that further refines the predicted pose by increasing its smoothness and offering joint torque estimates, which could be beneficial for healthcare and sports applications.

As a result, WheelPoser establishes a new state-of-the-art (SOTA) for sparse-IMU based upper body pose estimation of wheelchair users, achieving a mean joint angle error of $14.30^\circ$ and a mean joint position error of $6.74$~cm for real-time pose estimation. WheelPoser outperforms existing models designed for people without motor impairments by more than threefold.
To summarize, we make the following contributions:

\begin{itemize}
    \item We design and implement a first-of-its-kind real-time, sparse-IMU-based wheelchair user tracking system for practical and effective upper body pose estimation.
    \item We propose and verify an optimized IMU device setup, ensuring that it is minimally intrusive and practical for daily use while retaining essential information for accurate pose estimation.
    \item We collect and release a novel dataset that captures 167 minutes of essential on-wheelchair motions performed by 2 full-time wheelchair users and 12 participants without motor impairments.
    \item We comprehensively evaluate our system, establishing the first benchmark of sparse-IMU based upper body pose estimation for wheelchair users. The results demonstrate that our proposed method significantly enhances pose estimation accuracy compared to existing SOTA models.
    \item We discuss the broad application space enabled by our system and opportunities for future work.
\end{itemize}

In the sections below, we first briefly describe related work in motion tracking technologies for wheelchair users, followed by a description of WheelPoser's pose estimation pipeline. Next, we describe our data collection process for the WheelPoser-IMU dataset, our evaluation results, and some exemplary applications. We end by discussing the limitations of our system and potential avenues for future work.

\section{Related Work}
In this section, we review prior motion tracking systems with a focus on systems designed for wheelchair users.
We first look at general activity sensing technologies related to the movement of wheelchair users.
Then we review prior research on motion capture solutions for wheelchair users, including both camera-based and IMU-based solutions. Finally, we review pose estimation techniques that utilize sparse IMU measurements. For a more detailed review of motion capture technologies we refer readers to surveys by Desmarais et al.~\cite{desmarais2021review} and Nguyen et al.~\cite{nguyen2022review}

\subsection{Activity Sensing for Wheelchair Users}
Broadly in the field of motion tracking for wheelchair users, a substantial amount of research has been dedicated to tracking wheelchair kinematics and developing activity recognition techniques.
For instance, Rhodes et al.~\cite{rhodes2014validity} developed a radio-frequency based localization system for tracking the field positions of manual wheelchair sports players. 
Slikker et al.~\cite{van2015opportunities} explored the use of three wheelchair-mounted IMUs to measure motion features like speed and frame rotation during manual wheelchair sports. Similarly, SpokeSense~\cite{carrington2020spokesense} built a sensing system that could track manual wheelchair motion data (speed zone, orientation change) and provide sports related contextual information (dribbling sound, game buzzer) for adaptive basketball analytics. 
SMART\textsuperscript{wheel}~\cite{cooper2009smartwheel} is a commercial solution that can monitor wheelchair speed and other propulsion metrics like push force and frequency. G-WRM~\cite{hiremath2013development} employs a wheel-mounted gyroscope to measure angular velocities of the wheel, thereby allowing for speed measurements with known wheel size. De Vries et al.~\cite{de2023real} also developed a three-IMU based system that could track the linear velocity, traveled distance, and magnitude of turns of a wheelchair.

Prior research on tracking the physical activities of wheelchair users has primarily focused on fitness, with a substantial body of work dedicated to quantifying the energy expenditure of manual wheelchair users~\cite{grillon2017wireless,learmonth2016accelerometer,nightingale2015influence, hiremath2012predicting}. Additionally, researchers have investigated ways to track a set of discrete activities using cameras or IMUs, such as deskwork~\cite{garcia2015identifying}, household activities~\cite{hiremath2015detection, garcia2015identifying}, sitting posture~\cite{ma2016activity,ma2017activity}, wheelchair propulsion~\cite{french2008classifying,herrera2018towards,chen2018toward, de2023real, hiremath2015detection,garcia2015identifying}, and wheelchair transfers~\cite{wei2021automating,barbareschi2018use, garcia2015identifying}. More recently, commercially available fitness trackers - Apple Watch~\cite{WatchApp45:online} and Garmin~\cite{Wheelcha90:online} have integrated wheelchair settings to measure wheelchair users' calories burned, active minutes, number of pushes, etc.

WheelPoser advances beyond prior work by providing a more continuous and comprehensive user representation, rather than focusing solely on coarse-grained activities or wheelchair kinematics. This high-fidelity user representation could further have a trickle-down effect on higher-level techniques like activity recognition~\cite{Rajasegaran_2023_CVPRposeactivity, Choutas_2018_CVPRPotion, Shah_2022_WACVposeactivity2}.


\subsection{Motion Capture for Wheelchair Users with Cameras}
Most commonly, motion capture for wheelchair users relies on commercial marker-based motion capture systems, such as Vicon~\cite{ViconAwa72:online} and OptiTrack~\cite{OptiTrac63:online}, which have been the gold standard for both industrial and research applications due to their millimeter level accuracy.
For instance, these systems have enabled extensive research into the biomechanics of wheelchair propulsion~\cite{lin2004muscle, kukla2022symmetry,koontz2002shoulder, collinger2008shoulder} and transfer~\cite{nawoczenski2003three, kankipati2015upper, tsai2018upper}, as well as the pathology of various upper extremity injuries faced by wheelchair users~\cite{briley2020scapular, lewis2018injury, jayaraman2014shoulder}. The cost and setup requirements of these systems however often prohibit consumer use.

More recently, advancements in computer vision have led to the emergence of numerous markerless motion capture systems. RGB-based approaches have been developed to estimate both 2D~\cite{openpose, bazarevsky2020blazepose, yolov8pose, mmpose2020} and 3D~\cite{SMPL-X:2019, Kanazawa_2018_CVPRHMR, kocabas2020vibe, goel20234dhumans, ROMP} body pose with varying levels of granularity. However, these systems fail for wheelchair users, often leading to unnatural, incorrect pose estimates, due to a lack of representation in the training data.
To enhance the diversity of training data, researchers have recently explored generating synthetic image data using game engines for model refinement~\cite{huang2024wheelpose, black2023bedlam}, achieving varying levels of success.
Concurrently, depth-based systems have been employed by rehabilitation scientists to study the kinematics of wheelchair users.
Wei et al.~\cite{wei2016can} demonstrated that skeletal positions tracked by a Microsoft Kinect could be used to differentiate between proper and improper wheelchair transfer techniques. Similarly, Rammer et al.~\cite{rammer2018assessment} assessed the repeatability of the Kinect in extracting spatiotemporal parameters and joint range of motion during wheelchair propulsion and found high inter-trial repeatability. Furthermore, Milgrom et al.~\cite{milgrom2016reliability} examined the accuracy of the Kinect in tracking upper body joint angles during wheelchair propulsion compared to marker-based systems, finding discrepancies ranging from 6 to 26 degrees across different joints.


While camera-based approaches have shown increasing promise, they raise privacy concerns, are often sensitive to occlusion, lighting conditions, and user appearances (clothing, hair), and most importantly require a fixed sensor location and pose tracking space, making them impractical for on-the-go motion capture.

\subsection{Motion Capture for Wheelchair Users with Dense IMUs}
An alternative to camera-based systems is to instrument the user with sensors such as IMUs. Motion tracking using IMUs offers numerous advantages -- they are portable, robust to visual occlusion, work both indoors and outdoors, and are more privacy-preserving -- making them extremely suitable for everyday motion capture. 
Commercial IMU-based motion capture systems, such as Xsens~\cite{Xsens}, rely on the dense placement of IMUs (17 or more sensors) on each body joint to determine the orientations of these joints for pose estimation. Researchers have leveraged such setups to track wheelchair users' upper body kinematics during wheelchair transfer~\cite{kondo2024kinematic} and various functional activities~\cite{starrs2012biomechanical}. Similarly, Hooke et al.~\cite{hooke2009capturing} focused on tracking single-arm kinematics during wheelchair propulsion by placing three IMUs on the user's arm and one on their trunk.
However, the substantial quantity of sensors involved is intrusive for users and necessitates a time-consuming setup process, which significantly reduces the practicality and usability of such systems.

In contrast, WheelPoser employs a minimally intrusive setup with just four IMUs to achieve accurate upper body pose estimation. Our approach significantly enhances the practicality while retaining all the benefits of IMU-based solutions.

\subsection{Motion Capture with Sparse IMUs}
To improve the usability and practicality of IMU-based motion capture systems, researchers have recently started to investigate ways to leverage sparse inertial sensors for pose estimation.
Early work placed accelerometers on the wrists, trunk, and ankles~\cite{riaz2015motion, slyper2008action, tautges2011motion} and estimated pose by searching for similar accelerations in a prerecorded motion database.
SIP~\cite{von2017sparse} instead employed an optimization-based method to achieve offline full-body pose estimation with only 6 IMUs.
With the release of the large AMASS dataset (over 60 hours of motion sequences)~\cite{mahmood2019amass}, researchers started using deep learning based methods to predict body pose from IMU measurements. 
Specifically, DIP~\cite{huang2018deep} first adopted a bidirectional recurrent neural network (biRNN) to learn the mapping between IMU measurements and joint rotations from extensive motion capture datasets and showed a real-time human pose estimation with 6 IMUs. 
TransPose~\cite{yi2021transpose}, PIP~\cite{yi2022physical}, TIP~\cite{jiang2022transformer} and IMUPoser~\cite{mollyn2023imuposer} all built upon this by introducing new model architectures, prediction tasks and sensor locations to improve pose estimation. 
However, these systems are only designed for and evaluated with people without motor impairments. They all require placing IMU sensors on users' pelvis and legs, which is often challenging and uncomfortable for wheelchair users. Moreover, all existing systems are trained on motion capture datasets predominantly composed of standing and walking motions, which completely lack representation of the diverse range of wheelchair-related motions~\cite{cotton2020kinematic}.
Consequently, these limitations make them unusable for wheelchair users and ill-equipped to capture their poses accurately.  

In comparison, WheelPoser serves as the first-of-its-kind sparse-IMU-based motion capture system designed specifically for wheel\-chair users, from device setup and data synthesis to pose estimation. We also collect a novel motion capture dataset consisting of 167 minutes of on-wheelchair human motions and paired IMU data. This dataset is larger than the combined sizes of similar datasets in ambulatory settings (DIP~\cite{huang2018deep} and TotalCapture~\cite{trumble2017total}), and serves as an initial step towards creating inclusive motion capture datasets.

\section{WheelPoser System}

The primary design goals of our system are twofold: 1) to employ a sparse-IMU setup for enhanced usability and practicality, and 2) to achieve comprehensive upper body pose estimation in real-time to support a wide range of applications. This combination presents inherent challenges due to the ambiguity in mapping sparse inertial measurements to a broader set of joint angles. Although prior research has demonstrated the feasibility of using deep neural networks to learn this intricate mapping from large-scale motion capture datasets, there is currently no publicly available motion capture dataset of wheelchair users, let alone a large-scale dataset sufficient for deep neural network training.

We address these gaps and challenges from the following three perspectives: 1) we introduce a sparse-IMU configuration that captures the essential motions of both the user and the wheelchair while minimizing intrusiveness, 2) we propose a data synthesis and pose estimation pipeline that builds upon this device setup and enables us to leverage high-quantity and diverse upper body motions in existing datasets for wheelchair user pose estimation, and 3) we collect a novel motion capture dataset covering a wide range of essential wheelchair-related activities and fine-tune our pre-trained pose estimation model to bridge the gap between on-wheelchair activities and existing motion types.

\subsection{Device Setup}
\begin{figure}[h]
 \centering
\includegraphics[width=0.95\linewidth]{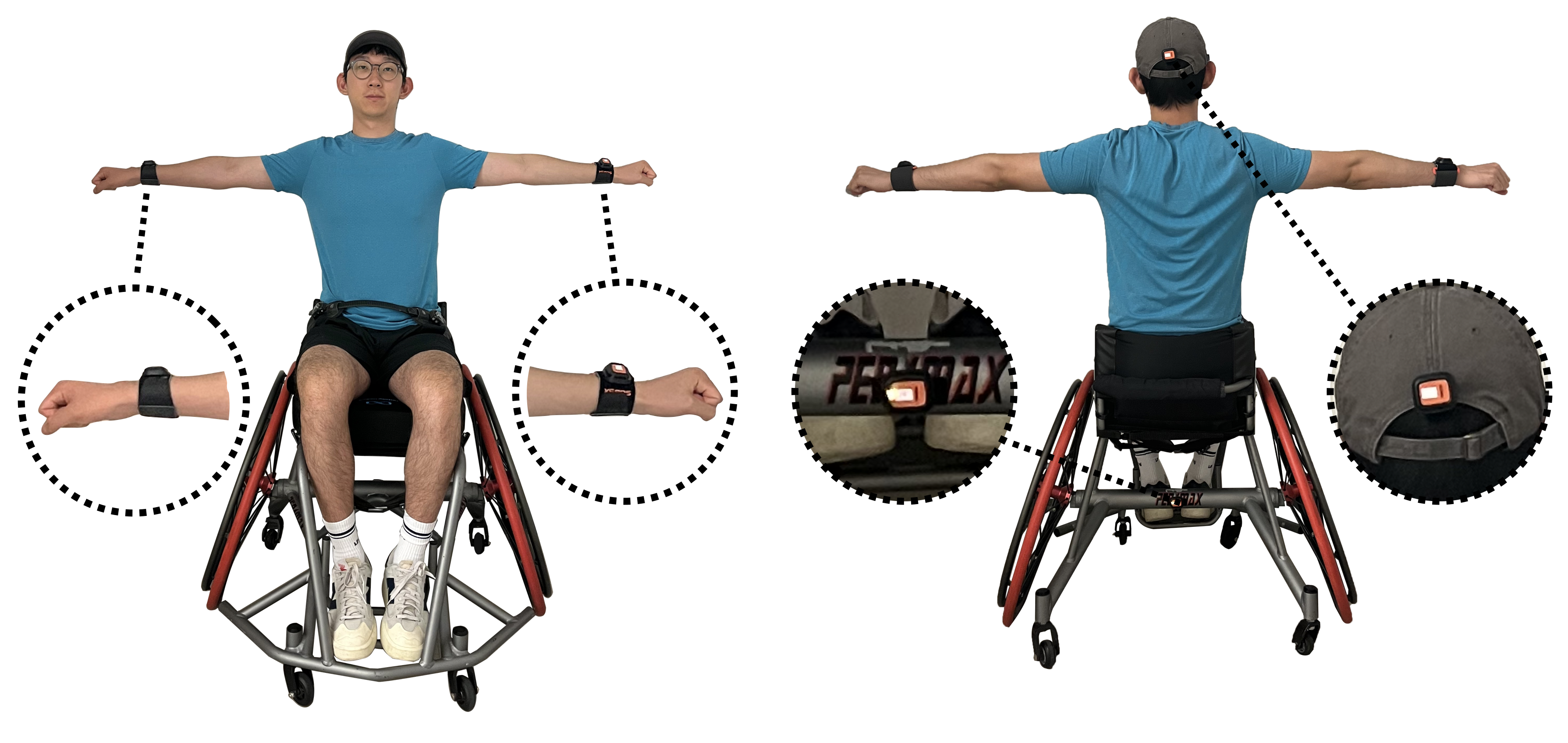}
\caption{WheelPoser uses four IMUs strategically placed on the user's forearms and head, as well as their wheelchair.}
\Description{This figure illustrates the device setup for WheelPoser, featuring two side-by-side photographs of the same person seated in a manual wheelchair. On the left, two IMUs are attached to the user's wrists. On the right, one IMU is attached to the back of the user's head, and another is attached to the wheelchair's axle.}
\label{fig:device}
\end{figure}
Our system comprises four IMUs strategically positioned on the user's body and their wheelchair to capture crucial aspects of the user's upper body motion (\autoref{fig:device}). Specifically, one IMU is placed on the user's head, primarily to capture the user's head movements but also certain upper body motions. The user's forearms are also instrumented with one IMU each to capture arm movements.
Traditionally, capturing the motion of the user's torso requires an additional IMU placed on their pelvis~\cite{huang2018deep, yi2021transpose, yi2022physical,jiang2022transformer}. However, for wheelchair users, we recognize that wearing a sensor on the pelvis could be not only inconvenient but also uncomfortable, considering their seated posture and the presence of wheelchair backrests for trunk support. Therefore, we propose to position this sensor at the center of the wheelchair axle, which is a standard component in modern wheelchairs. Our intuition is that as modern wheelchairs are designed to provide wheelchair users with spinal stabilization, the wheelchair's motion will exhibit strong correlations with the user's pelvis motion.
To validate our intuition, we conducted a preliminary study.

\subsubsection{Preliminary Study: Examining Wheelchair-Pelvis Motion Correlation}
To examine our intuition, we conducted a preliminary study with three participants without motor impairments (two males and one female) sitting on a provided manual wheelchair. The participants were asked to perform a diverse set of typical wheelchair motions, such as wheelchair propulsion and turning (Sec.~\ref{collection_protocol}). 
In addition to the four IMUs of our device setup, we attached the fifth IMU to the user's pelvis to capture the actual pelvis motion for comparison.
All five IMUs were time-synchronized and streamed at 30Hz, resulting in a total of 100 minutes of motion data.

Our analysis focused on two main aspects. First, we examined the similarity of orientations and accelerations between the user's pelvis and wheelchair. Second, interrelated with our later discussed data synthesis and pose estimation pipeline (Sec.~\ref{data_syn}), we investigated the correlation between the relative accelerations of upper body joints (specifically the wrists and head) in relation to the wheelchair and the user's pelvis. For ease of reference, we term this type of acceleration as \textit{"normalized acceleration"}.
Additionally, all IMU signals were transformed from the sensor-coordinate frame to a body-centric frame for analysis, where the x-axis points to the left, the y-axis points upward, and the z-axis points forward, ensuring consistency with the human kinematic model used in our study (Sec.~\ref{data_syn}).

\begin{figure*}[h]
 \centering
\includegraphics[width=0.75\textwidth]{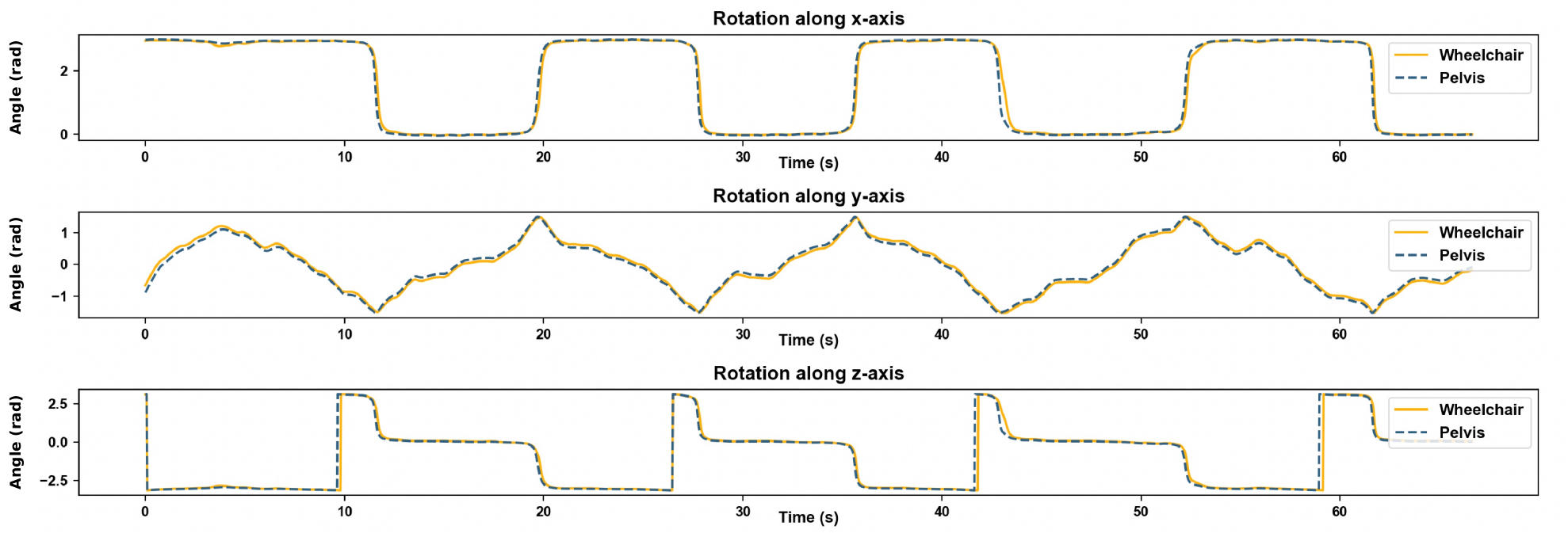}
\caption{Sample orientation measurements (Euler angles) of two IMUs attached to the user's pelvis and the wheelchair's central axle. The orientations are highly correlated, affirming the close relationship between the pelvis and wheelchair axle motion.}
\Description{This figure presents three line charts of orientation signals from two IMUs. In each chart, two series of signals are displayed: one from the IMU attached to the user’s pelvis and another from the IMU attached to the wheelchair. The figure demonstrates that the orientation signals from the two IMUs are closely aligned, indicating a strong correlation.}
\label{fig:ori_pilot}
\end{figure*}

\begin{figure*}[h!]
 \centering
\includegraphics[width=0.8\textwidth]{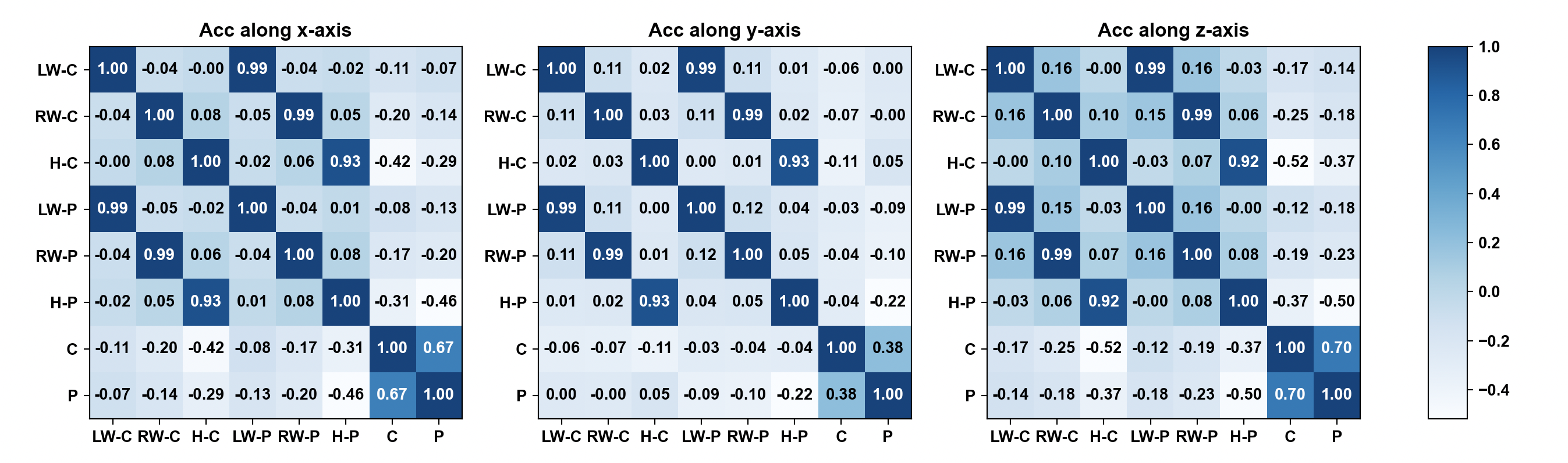}
\caption{Correlation coefficient matrices displaying normalized accelerations of the left wrist (LW), right wrist (RW), head (H), pelvis (P), and wheelchair (C), where, for instance, LW-C represents the normalized acceleration of the left wrist with respect to the wheelchair. normalized accelerations with respect to the pelvis and wheelchair are highly correlated.}
\Description{This figure displays the correlation coefficient matrix for eight different acceleration signals. From top to bottom, they include: normalized left wrist acceleration relative to the wheelchair, normalized right wrist acceleration relative to the wheelchair, normalized head acceleration relative to the wheelchair, normalized left wrist acceleration relative to the pelvis, normalized right wrist acceleration relative to the pelvis, normalized head acceleration relative to the pelvis, acceleration of the wheelchair, and acceleration of the pelvis. There are three matrices depicting correlations along the x, y, and z axes, respectively. The figure indicates that the normalized accelerations are strongly correlated, and the accelerations of the wheelchair and the pelvis are also closely correlated.}
\label{fig:acc_pilot}
\end{figure*}

\textit{Orientation:}
To measure the similarities between the orientations of the user's pelvis and the wheelchair, we used the distance metric (\autoref{equation:ori}) proposed by Huynh~\cite{huynh2009metrics} where $q_1$ and $q_2$ are unit quaternions, and the function $\phi$ gives values in the range $[0,1]$, with 0 representing that the two quaternions are identical. Using this metric, the average distance between the orientation of the user's pelvis and the orientation of the wheelchair is 0.007 (std = 0.011), indicating that the two orientations are indeed quite similar. 
We also qualitatively examined this similarity by using their Euler angle representations and found consistent correlations across all samples (\autoref{fig:ori_pilot}).
\begin{equation}
\phi(q_1,q_2) = 1 - |q_1 \cdot q_2|
\label{equation:ori}
\end{equation}

\textit{Acceleration:}
For accelerations, we assessed similarity by calculating correlation coefficients along each axis (ranging from 0 to 1, with 1 indicating identical signals). 
Here and throughout the rest of this paper, we note that "acceleration" refers specifically to gravity-subtracted linear acceleration.
The resulting coefficient matrices for each axis are illustrated in Figure~\ref{fig:acc_pilot}. Specifically, the coefficients between the user's pelvis acceleration and the wheelchair's acceleration are 0.67, 0.38, and 0.70, respectively, indicating a fairly strong correlation. The lower coefficients along the y-axis can be attributed to noise, as there is minimal motion along the axis perpendicular to the ground. For the normalized accelerations of the user's wrists and head with respect to the user's pelvis and the wheelchair, the average coefficients for the left wrist, right wrist, and head are 0.989, 0.99, and 0.926, respectively, which demonstrate strong correlations for pose estimation.

\textit{Summary: }
Based on this preliminary study, we have confirmed the correlation between the motion of the user's pelvis and that of the wheelchair, validating the effectiveness of our device setup for capturing the upper body pose of wheelchair users. Additionally, the correlations observed between the normalized accelerations hold promise for leveraging diverse upper body motions in existing datasets collected from individuals without motor impairments. Furthermore, we envision that this configuration could be seamlessly integrated with devices that users already carry on a daily basis, such as smartwatches and earbuds, as well as with low-cost devices in the chairable form factor~\cite{carrington2014wearables, mollyn2023imuposer}. 

\subsection{Learning-based Kinematics Module}
WheelPoser uses a deep neural network to predict upper body joint angles from sparse-IMU measurements. We propose a data synthesis and pose estimation pipeline that enables us
to first train a base model with a broad understanding of upper body pose patterns using diverse upper body motions found in existing datasets and later fine-tune this model with a small amount of data to bridge the gap between existing motions and wheelchair mobility. Here we describe our data synthesis pipeline followed by the architectures of three neural networks we adapt from SOTA pose estimation models using sparse-IMUs.

\subsubsection{Data Synthesis} \label{data_syn}
To train the neural networks in our kinematics module, we need a sufficiently large dataset that contains both IMU measurements and human pose ground truths.
Following previous work, we adopt the SMPL body model~\cite{smpl} as our kinematic model and the AMASS dataset~\cite{mahmood2019amass} to train our base model. The AMASS dataset is a composition of existing motion capture datasets and contains SMPL pose parameters for more than 60 hours of motions performed by over 300 subjects without motor impairments. However, as the AMASS dataset does not contain IMU sensor data, we synthesize inertial data by attaching virtual IMUs to the corresponding vertices of our selected joints on the SMPL mesh, similar to DIP~\cite{huang2018deep}. We then calculate the sequential positions, accelerations, and rotations of these vertices in the global reference frame using the SMPL body model (\autoref{equation:acc}). 
Here, $x_i(t)$ and $a_i(t)$ represent the position and the acceleration of the $i$ th IMU at the frame $t$ respectively, and $\Delta t$ represents the time interval between two adjacent frames which is $\frac{1}{60}s$, as we set the frame rate to 60~Hz. Following prior work, we set $n=4$ to mitigate motion jitters in the data.

\begin{equation}
 a_i(t) = \frac{x_i(t-n)+x_i(t+n)-2x_i(t)}{(n\Delta t)^2}, i = 1, 2, 3, 4
 \label{equation:acc}
\end{equation}

Informed by our preliminary study, we also normalize the synthetic upper body IMU data with respect to the pelvis joint. Specifically, we have the synthetic accelerations
$[a_{pelvis}, \ a_{larm}, \ a_{rarm}, \\ \ a_{head}] \in \mathbb{R}^{3\times4}$ and rotations $[R_{pelvis}, \ R_{larm}, \ R_{rarm}, \ R_{head}] \in \mathbb{R}^{3\times3\times4}$ in the same global frame aligned with the SMPL reference frame. We normalize the inertial measurements of the leaf joints (left arm, right arm, and head) with respect to the pelvis joint as:
\begin{equation}
    \tilde{a}_{leaf} = R^{-1}_{pelvis}(a_{leaf} - a_{pelvis})
\end{equation}
\begin{equation}
    \tilde{R}_{leaf} = R^{-1}_{pelvis}R_{leaf}
\end{equation}
and the inertial measurements of the pelvis joint as:
\begin{equation}
    \tilde{a}_{pelvis} = R^{-1}_{pelvis}a_{pelvis}
\end{equation}
\begin{equation}
\tilde{R}_{pelvis} = R_{pelvis}
\end{equation}

\subsubsection{Kinematics Module Network Architecture}
The input to our kinematics module is a concatenated vector of size 48, which includes the normalized accelerations $[\tilde{a}_{pelvis}, \ \tilde{a}_{larm}, \ \tilde{a}_{rarm}, \ \tilde{a}_{head}] \\ \in \mathbb{R}^{3\times4}$, and normalized orientations expressed in rotation matrix format $[\tilde{R}_{pelvis}, \ \tilde{R}_{larm}, \ \tilde{R}_{rarm}, \ \tilde{R}_{head}] \in \mathbb{R}^{3\times3\times4}$, captured from the four IMUs.
The output of our kinematics module is a vector of 96 SMPL pose parameters - 16 upper body joints represented as 6D rotations~\cite{zhou2019continuity}.
Here, unlike previous approaches that directly measure pelvis rotation using the IMU placed on the pelvis~\cite{huang2018deep, yi2021transpose, yi2022physical, jiang2022transformer}, we opt to predict the pelvis pose in the global reference frame. This approach compensates for discrepancies between the orientations of the user’s pelvis and the wheelchair, offering a more accurate representation of the movement dynamics of wheelchair users.
Specifically, we propose three candidate network architectures for our kinematics module adapted from existing work on sparse-IMU-based pose estimation:

\begin{itemize}
    \item \textbf{Single-stage biRNN:} Inspired by DIP~\cite{huang2018deep} and IMUPoser~\cite{mollyn2023imuposer}, this architecture contains a two-layer bidirectional recurrent neural network (biRNN) with long short-term memory (LSTM) cells of hidden dimension 256. We first transform the input vector to this hidden dimension with a ReLU-activated linear layer. Next, we feed these embeddings sequentially into the LSTM after which they are transformed linearly into estimated SMPL pose parameters.
    \item \textbf{Three-stage biRNN:} Previous research has demonstrated that utilizing joint positions as an intermediate representation greatly enhances a model’s ability to learn complex human motion priors. The second network in our study accordingly adopts a three-stage structure, inspired by TransPose~\cite{yi2021transpose}. Here, each stage features a biRNN with a structure similar to the single-stage biRNN previously described. The first stage predicts the positions of the head, left wrist, and right wrist joints based on the input vector. Subsequently, the second stage utilizes both the initial input vector and the output from the first stage to predict all upper body joint positions. The final stage then uses both the input vector and the output from the second stage to predict the SMPL pose parameters. The LSTM cells in each stage have hidden dimensions of 256, 64, and 128, respectively.
    \item \textbf{Transformer decoder:} We also compare against using a transformer-based model for pose estimation. We follow a similar structure to TIP~\cite{jiang2022transformer} and use an autoregressive transformer decoder (4 layers, 16 heads, model dimension 256) followed by a 1-layer RNN to estimate SMPL pose parameters. Different from TIP, we use a physics-based optimization module~(Sec.~\ref{sec:physics-based-optim}) and do not estimate stable body points or root velocity.
\end{itemize} 

\subsection{Physics-based Optimization Module}
\label{sec:physics-based-optim}
So far, our pose estimation pipeline has only considered kinematic information neglecting human motion dynamics, which can result in unnatural and jittery motion in real-time pose estimation as shown in previous work~\cite{yi2021transpose, yi2022physical, jiang2022transformer}. Additionally, in light of the potential applications of our system, such as health monitoring and sports analytics, where access to information about the joint torque of wheelchair users' upper body can be extremely helpful, we have decided to incorporate a physics-based optimization module to address both of these aspects without adding any additional sensors into our system.
\begin{equation}
    \ddot{\theta}_{ref}=k_{p}(\theta_{k}\ -\ \theta) \ - \ k_{d}\dot{\theta}
    \label{equation:pd_controller}
\end{equation}

Inspired by PIP~\cite{yi2022physical}, our physics-based optimization module first employs a joint rotation proportional-differential (PD) controller to compute the reference joint angular acceleration $\ddot{\theta}_{ref}$ from the joint angles $\theta_{k}$ estimated by the learning-based kinematics module (\autoref{equation:pd_controller}, $\theta$ and $\dot{\theta}$ are the joint angles and angular velocities, respectively).
The gain parameters $k_{p}$ and $k_{d}$ are set to 3600 and 60, respectively, following prior work~\cite{yi2022physical}.
These reference accelerations are then refined using quadratic programming to comply with a set of physical constraints, including the equation of motion~\cite{featherstone2014rigid} as well as no-sliding restrictions for all contact points between the user and the environment. For a more detailed explanation of the quadratic programming problem, we refer readers to PIP~\cite{yi2022physical} and PhysCap~\cite{shimada2020physcap}.
Here, unlike previous approaches that primarily model foot-ground contacts during walking, we focus on the contact between the user and the wheelchair, defining contact joints as the left hip, right hip, left foot, and right foot based on common wheelchair sitting scenarios. Moreover, considering that user-wheelchair contact points move together, we optimize the pose in the pelvis-relative frame rather than the global frame used in the existing work. Finally, we perform double integration on the refined joint angular accelerations to derive the optimized pose.

\begin{figure}[h!]
 \centering
\includegraphics[width=0.99\linewidth]{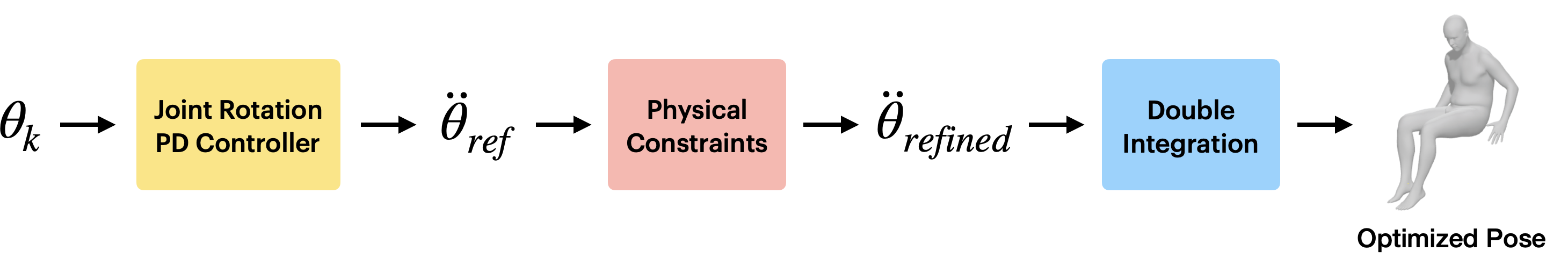}
\caption{Overview of the physics-based optimization module: It calculates reference joint angular acceleration from the kinematics module's estimates. These accelerations are then refined with physical constraints and double-integrated to produce optimized poses.}
\Description{This figure illustrates the detailed components of the physics-based optimization module. It includes a joint rotation PD controller that calculates the reference joint angular acceleration based on the joint angles estimated by the kinematics module. The reference accelerations are then refined using a set of physical constraints and finally double-integrated to obtain the optimized poses.}
\label{fig:physics}
\end{figure}

\section{WheelPoser-IMU Dataset}
\begin{table*}[t]
\begin{tabular}{|c|c|c|c|c|c|c|c|}
\rowcolor[HTML]{EFEFEF}
\hline
ID  & Gender & Age & Height (cm) & Arm length (cm) & Time Using Wheelchair&Wheelchair Model\\ \hline
W1 & Male   & 48  & 179            & 58.4 &27 years & TiLite Aero X             \\ 
W2 & Male   & 44  & 183            & 61.1 &12 years & TiLite Aero T               \\ 
\hline
\end{tabular}
\vspace*{2mm}
\caption{Demographic information of full-time wheelchair user participants.}
\label{tab:demographics_2}
\end{table*}
As there is no existing motion capture dataset that encompasses typical movements performed by wheelchair users, we sought to collect our own. Our dataset, WheelPoser-IMU, contains 167 minutes of paired motion capture and IMU sensor data from 14 participants (including 2 full-time wheelchair users and 12 participants without motor impairments) performing various on-wheelchair daily activities. Here we describe our data collection apparatus, procedure, participant demographics, and data processing pipeline.

\subsection{Data Collection Apparatus}
\begin{figure}[b]
 \centering
\includegraphics[width=0.9\linewidth]{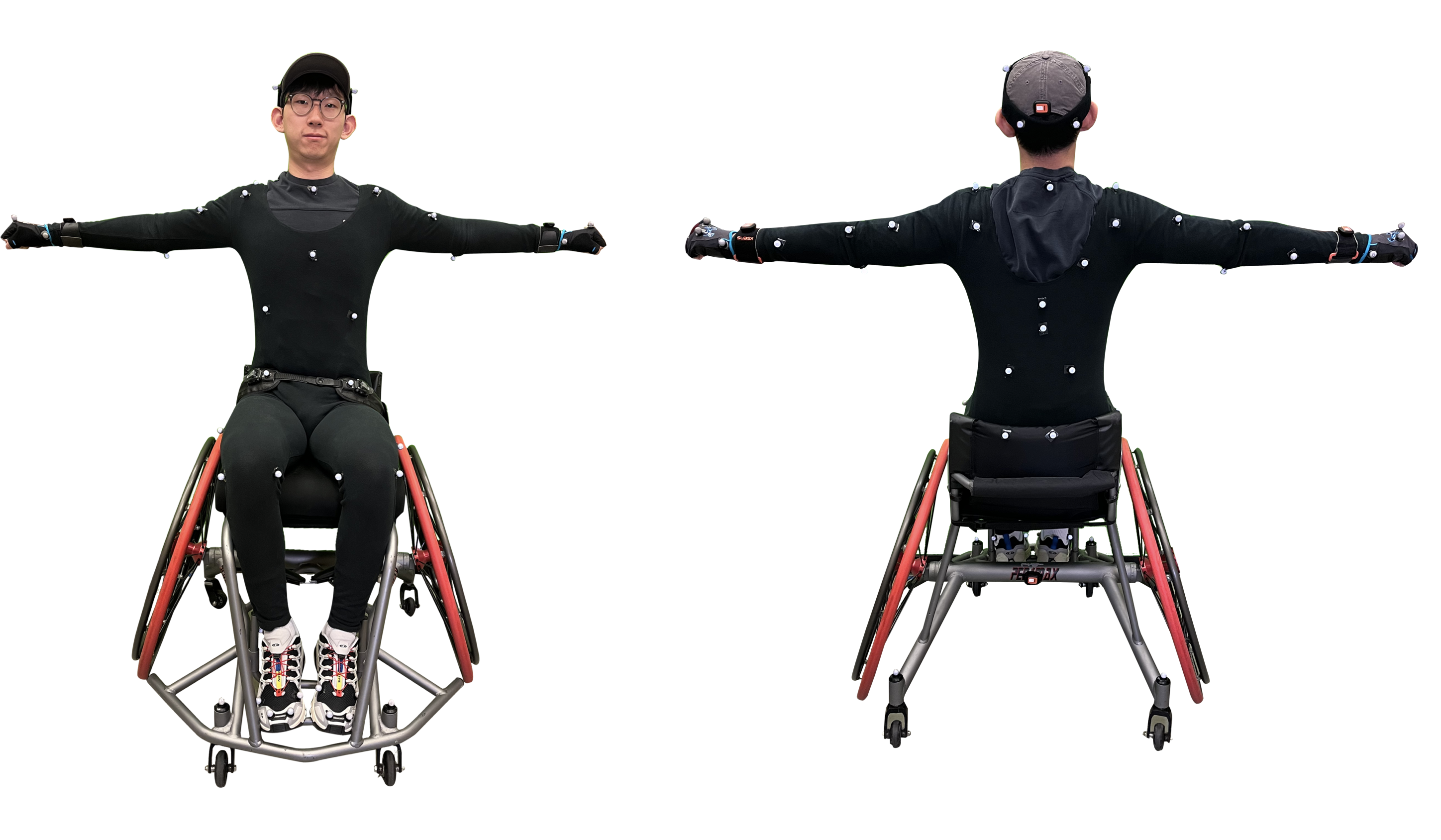}
\caption{The WheelPoser-IMU data collection setup includes four IMUs placed on both the participant and their wheelchair, along with 60 retroreflective markers attached across bony landmarks on the participant's body.}
\Description{This figure illustrates the device setup for data collection, featuring two side-by-side photographs of the same individual seated in a manual wheelchair. On the left, two IMUs are attached to the user's wrists. On the right, one IMU is attached to the back of the user's head, and another is affixed to the wheelchair's axle. Additionally, a total of 60 markers are attached to the user's body, distributed around important bony landmarks, such as the wrists and shoulder joints.}
\label{fig:marker_set}
\end{figure}

Our data collection setup comprised four Movella DOTs~\cite{MovellaD68:online}, each configured to sample data at a rate of 60 Hz, which is the maximum sampling rate supported for real-time streaming. The four Movella DOTs were time-synchronized and communicated over Bluetooth to a nearby laptop for recording and processing. To capture ground truth pose, we employed a Vicon Motion Capture System consisting of twelve MX40 cameras and four T160 cameras, operating at a sampling rate of 120 FPS. The participants were instrumented with 60 retroreflective markers (Figure~\ref{fig:marker_set}). The Vicon motion capture data was down-sampled and synchronized with the concurrently collected IMU data streams. Following prior work~\cite{huang2018deep}, we further fit an SMPL mesh to this motion capture data using Mosh++~\cite{mahmood2019amass}. 

\subsection{Participants and Procedure}\label{collection_protocol}
\begin{table}[b]
\renewcommand{\arraystretch}{1.1}
\begin{tabular}{|c|c|c|c|c|}
\rowcolor[HTML]{EFEFEF}
\hline
ID  & Gender & Age & Height (cm) & Arm length (cm) \\ \hline
P1  & Male   & 28  & 175            & 56.7               \\ 
P2  & Male   & 28  & 180            & 58.3                \\ 
P3  & Female & 27  & 164            & 54.2                \\ 
P4  & Male   & 24  & 174            & 58.5                \\ 
P5  & Male   & 23  & 183            & 61.6               \\ 
P6  & Female & 25  & 168            & 53.9               \\ 
P7  & Female & 27  & 165            & 54.9               \\ 
P8  & Male   & 32  & 173            & 56.7                \\ 
P9  & Female & 28  & 160            & 51.1                \\ 
P10 & Male   & 24  & 172            & 54.7                \\ 
P11 & Female & 27  & 161             & 52.8                \\ 
P12 & Female & 27  & 160            & 52.3                \\ 
\hline
\end{tabular}
\vspace*{2mm}
\caption{Demographic information of participants without motor impairments.}
\label{tab:demographics_12}
\end{table}

We recruited 14 participants in total for our data collection study, including 2 full-time manual wheelchair users and 12 participants without motor impairments.
The two full-time wheelchair user participants had been using a wheelchair for 27 and 12 years, respectively. Detailed demographics information are show in Table~\ref{tab:demographics_2} and ~\ref{tab:demographics_12}. Participants first received instructions on various wheelchair movement techniques and were then allowed to familiarize themselves with these motions until they felt confident. Practice times ranged from 16 minutes to 50 minutes. Full-time wheelchair user participants were asked to use their own manual wheelchairs (TiLite Aero T and Aero X) for data collection to ensure comfort and external validity, while other participants were provided with a PER4MAX Thunder lightweight manual wheelchair.
Before starting our study, we performed a vendor-recommended magnetic field calibration and heading reset procedure on the four IMUs to account for magnetic offsets. Following this, the IMU sensors were attached to the participant and their wheelchair (\autoref{fig:device}). We then performed a calibration process to map the IMU measurements to a body-centric frame and to compensate for the offset between the sensor and the corresponding bone. For further details on calibration, we refer readers to~\cite{huang2018deep} and~\cite{yi2021transpose}.

During our study, we collected the following typical motions that wheelchair users would perform on a daily basis, following established wheelchair skill testing procedures~\cite{lindquist2010reliability}:
\begin{itemize}
  \item \textbf{Arm Motion:}  Left arm raises (0, 45, 90 degrees), left arm raising overhead, left arm crossing the torso, right arm raises (0, 45, 90 degrees), right arm raising overhead, right arm crossing the torso, both arms raises (0, 45, 90 degrees), both arms raising overhead, both arms crossing the torso, both arms crossing behind the head, both arms swinging.
  \item \textbf{Upper Body:} Rotation to the left, rotation to the right, Leaning forward, leaning to the left, leaning to the right, leaning diagonally to the left, leaning diagonally to the right.
  \item \textbf{Wheelchair Locomotion:} Push forward, push backward, push forward in a circle, push backward in a circle, turn while moving forward, turn-in-place, pivot turn.
\end{itemize}

Participants performed these motions in a continuous manner within a recording session (e.g. arm motions) in order to capture natural transitions between categories. Before the start of each recording session, participants were asked to perform five rapid arm raises (the upper body motion seen in typical jumping jacks), to help synchronize motion capture data with IMU sensor data. Data collection typically took 45 minutes for participants without motor impairments and 1 hour for wheelchair user participants.
We note that this 15-minute difference was primarily due to longer rest periods between sessions for wheelchair user participants. We also encountered more instances of mocap markers falling off for these participants, requiring extra time to reposition. The total duration of collected motions was consistent across all participants.
Compensation of \$15 and \$75 was provided to each participant without motor impairments and wheelchair user participant, respectively.


\subsection{Data Processing}
The collected IMU data was first interpolated for missing packets and later synchronized with the SMPL mesh data based on the acceleration markers. In addition to the calibration performed at the beginning of the data collection, the IMU data for each individual session were further calibrated based on the corresponding joint orientation of the first frame of the SMPL mesh to compensate for drift errors that occurred during the data collection process. Later, the arm-raising motions at the beginning of each session were discarded, resulting in an overall 167 minutes of synchronized IMU and SMPL data on various wheelchair-use-related motions, which is nearly twice the size of previously collected ambulatory motion dataset (DIP-IMU and TotalCapture). 

\section{Evaluation}\label{Eval}

In this section, we first outline the dataset and metrics used for the model training and evaluation in Sec.~\ref{metrics}. Leveraging these datasets and metrics, we then quantitatively evaluate the performance of SOTA models designed for people without motor impairments and our proposed kinematics modules in Sec.~\ref{kinematics}. Using the kinematics module that achieves the best results, we evaluate the effectiveness of our proposed physics-based optimization module in Sec.~\ref{physics}. Lastly, we delve into a detailed exploration of the model's performance across various on-wheelchair motion types in Sec.~\ref{types}.

All training and evaluation processes were conducted on a computer with an Intel(R) Core(TM) i7-13700KF CPU and an NVIDIA RTX 4080 graphics
card. The kinematics modules were trained with a batch size of 256 using the Adam optimizer and a learning rate of $0.001$. 
We trained all kinematics modules using the mean squared error (MSE) loss function. 
In line with previous work, a sliding window of 26 IMU measurement frames was used for real-time pose estimation. Specifically, for biRNN-based modules, the window includes 20 past frames, 1 current frame, and 5 future frames, whereas for the transformer-based module, it comprises 25 past frames and 1 current frame.

\subsection{Dataset and Metrics}\label{metrics}
\begin{figure}[t]
    \centering
    \includegraphics[width=\linewidth]{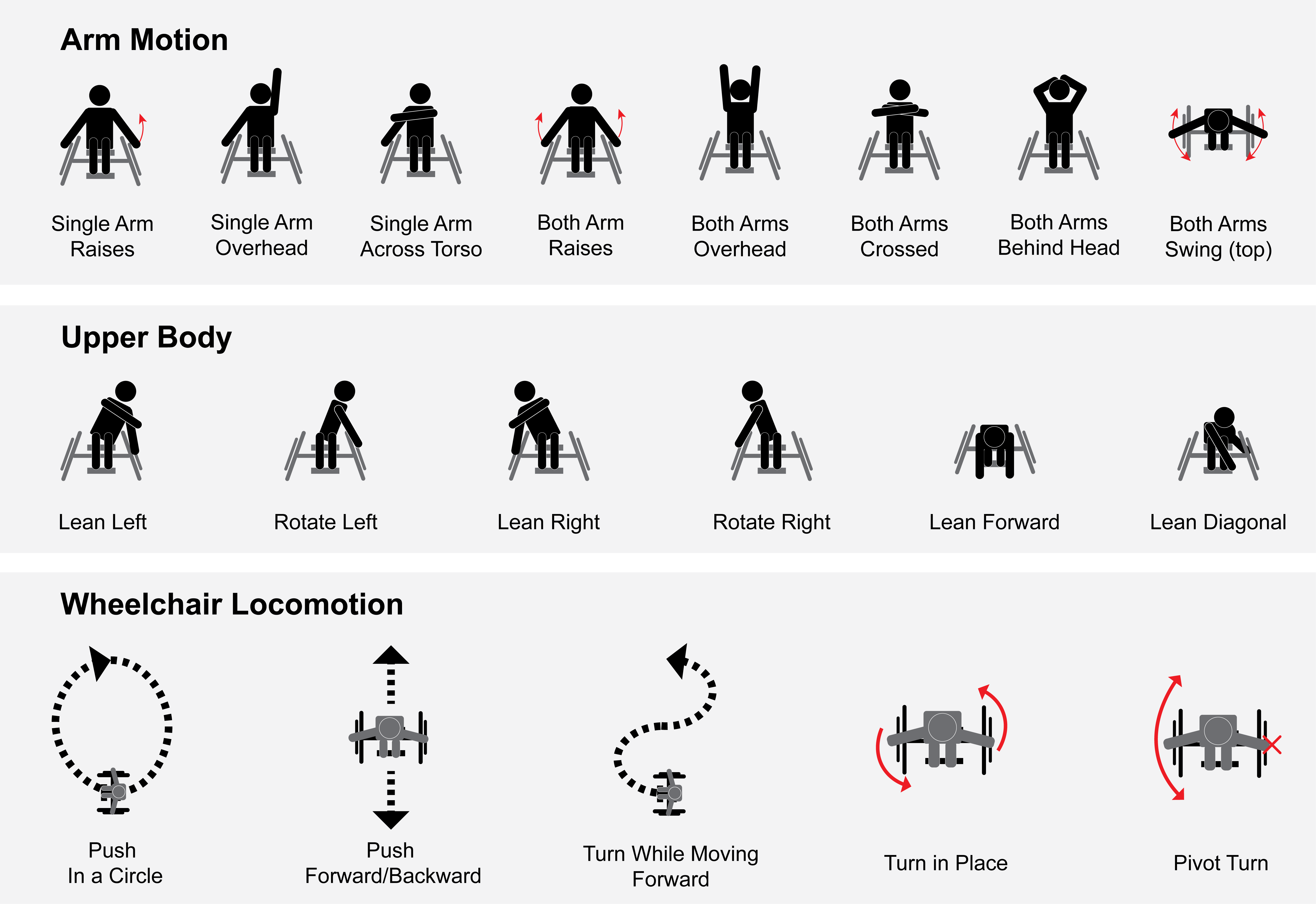}
    \caption{Exemplary motion sequences and poses collected in the WheelPoser-IMU dataset.}
    \Description{This figure showcases the motions collected in the WheelPoser-IMU dataset. Three rows of icons are displayed. The top row depicts the arm motions captured, including the left arm raised at 0, 45, and 90 degrees, left arm overhead raise, left arm across the torso, both arms raised at 0, 45, and 90 degrees, both arms across the torso, both arms crossed behind the head, and both arms swinging. The second row presents various upper body motions, such as leaning to the left, rotating to the left, leaning to the right, rotating to the right, leaning forward, and leaning diagonally to the left and right. The third row illustrates several wheelchair locomotions, including pushing forward and backward, pushing in a circle, turning while moving forward, turning in place, and executing pivot turns.}
    \label{fig:poses}
\end{figure}
For training the kinematics modules, we first utilize data synthesized from AMASS, followed by fine-tuning using our collected WheelPoser-IMU dataset.
To assess the effectiveness and external validity of our proposed method, we conducted a leave-one-subject-out evaluation study exclusively on the full-time wheelchair users' data. Specifically, we upsample data from a single wheelchair user and combine it with data from 12 participants without motor impairments to create a fine-tuning dataset, of which 40\% comprises the single full-time wheelchair user's data. The fine-tuned model is then evaluated on the excluded wheelchair user's data, and we report the averaged results from two full-time wheelchair users. 
We use the following commonly used metrics to quantitatively evaluate estimated poses:

\begin{enumerate}
  \item \textit{Joint angular error}, which measures the mean global rotation error of estimated upper body joints in degrees.
  \item \textit{Joint position error}, which measures the mean Euclidean distance error of estimated upper body joints in centimeters with the pelvis joint aligned.
  \item \textit{Mesh error}, which measures the mean Euclidean distance error of all vertices of the estimated upper body mesh also with the pelvis joint aligned and body model in mean shape
  \item \textit{Jitter error}, which measures the average jerk of estimated upper body joints, which is the third derivative of position with respect to time and reflects the naturalness of the motion.
\end{enumerate}

Moreover, we incorporate wrist joint position error and elbow joint position error as two additional metrics. This stems from the significance of wrist position in determining the propulsion pattern of wheelchair users~\cite{boninger2002propulsion}, while elbow positions provide valuable insights into the upper extremity's range of motion and the associated risk for overhead reaching~\cite{li2023breaking}.
To obtain the estimated joint positions and body mesh, we used the mean SMPL shape parameters, ensuring consistency with prior work and mimicking real-world scenarios where users' specific shape parameters are typically unavailable.

\begin{figure*}[b]
     \centering
     \begin{subfigure}[t]{0.22\textwidth}
         \centering
         \includegraphics[width=\textwidth]{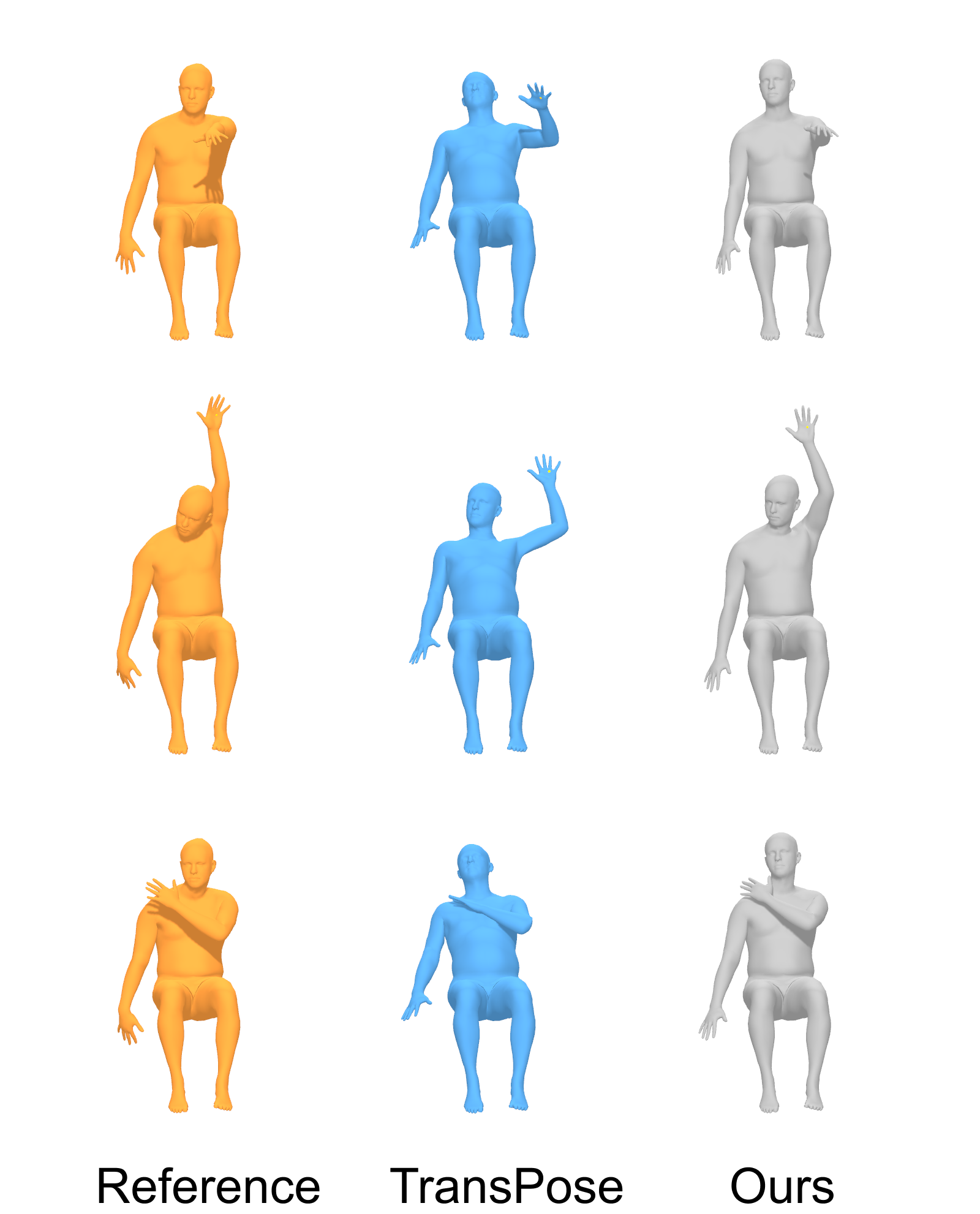}
         \caption{Arm reaching motion}
         \label{fig:qual_arm}
     \end{subfigure}
     \hfill
     \begin{subfigure}[t]{0.22\textwidth}
         \centering
         \includegraphics[width=\textwidth]{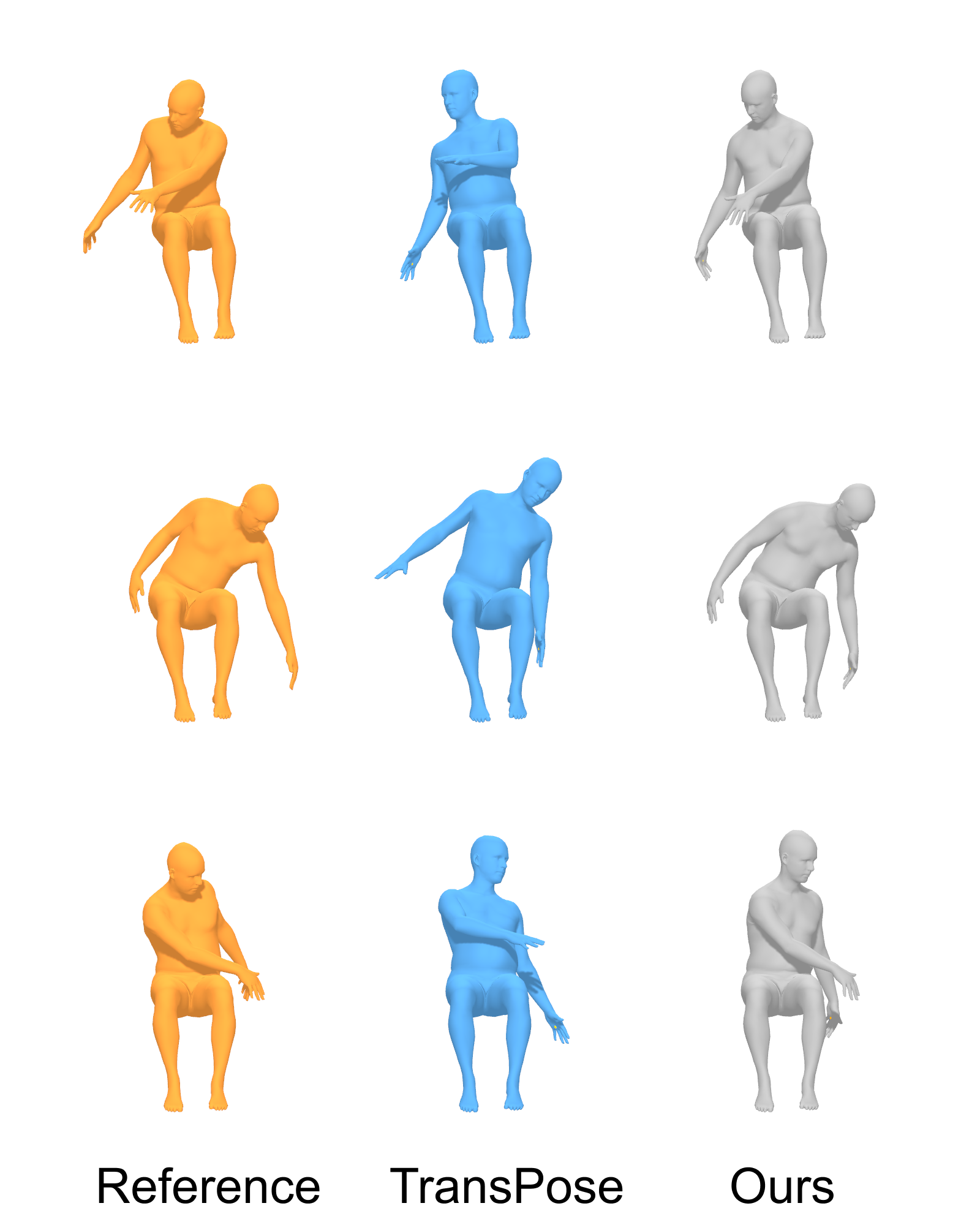}
         \caption{Torso rotation.}
         \label{fig:qual_body}
     \end{subfigure}
     \hfill
     \begin{subfigure}[t]{0.46\textwidth}
         \centering
         \includegraphics[width=0.48\linewidth]{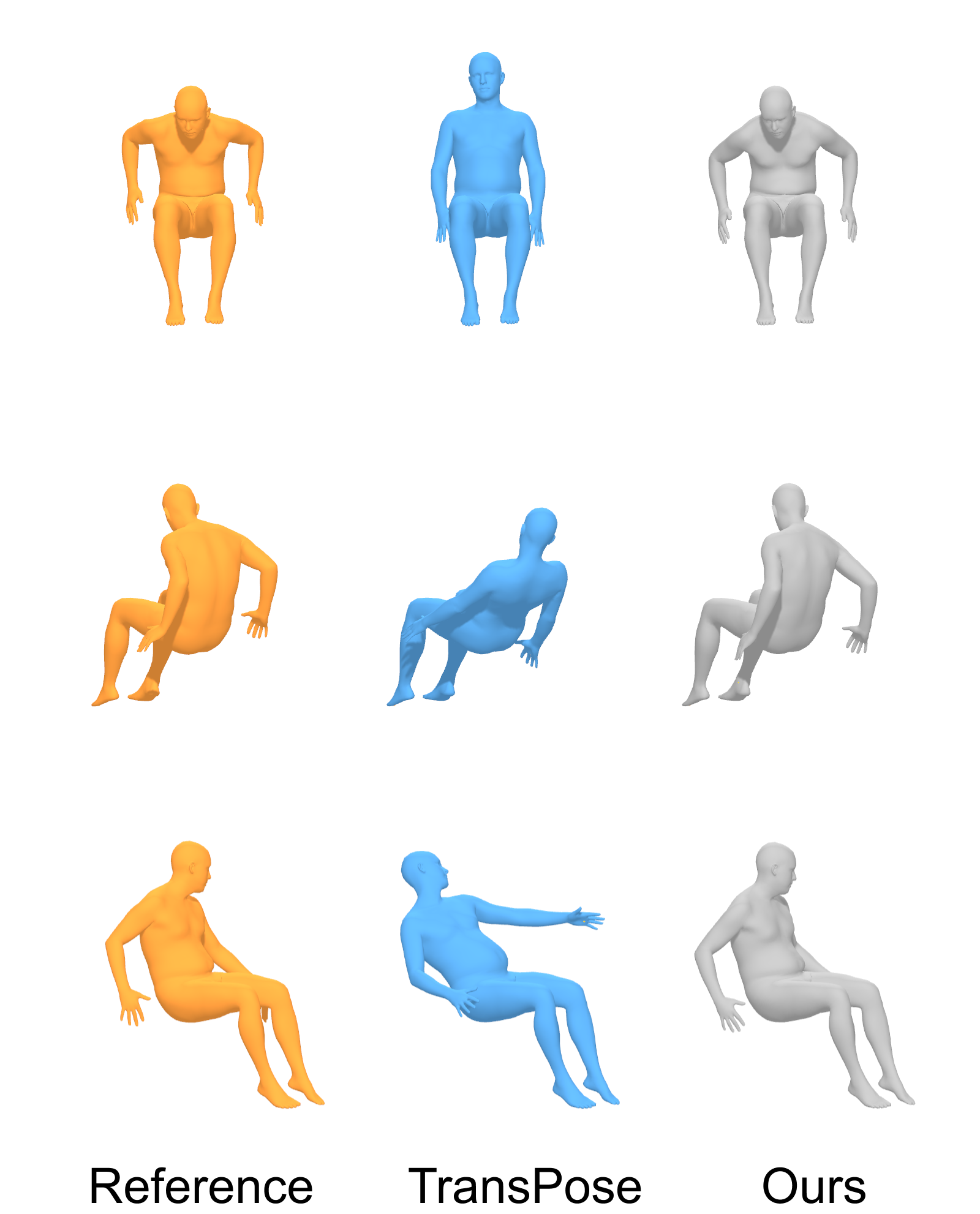}
         \hfill
        \includegraphics[width=0.48\linewidth]{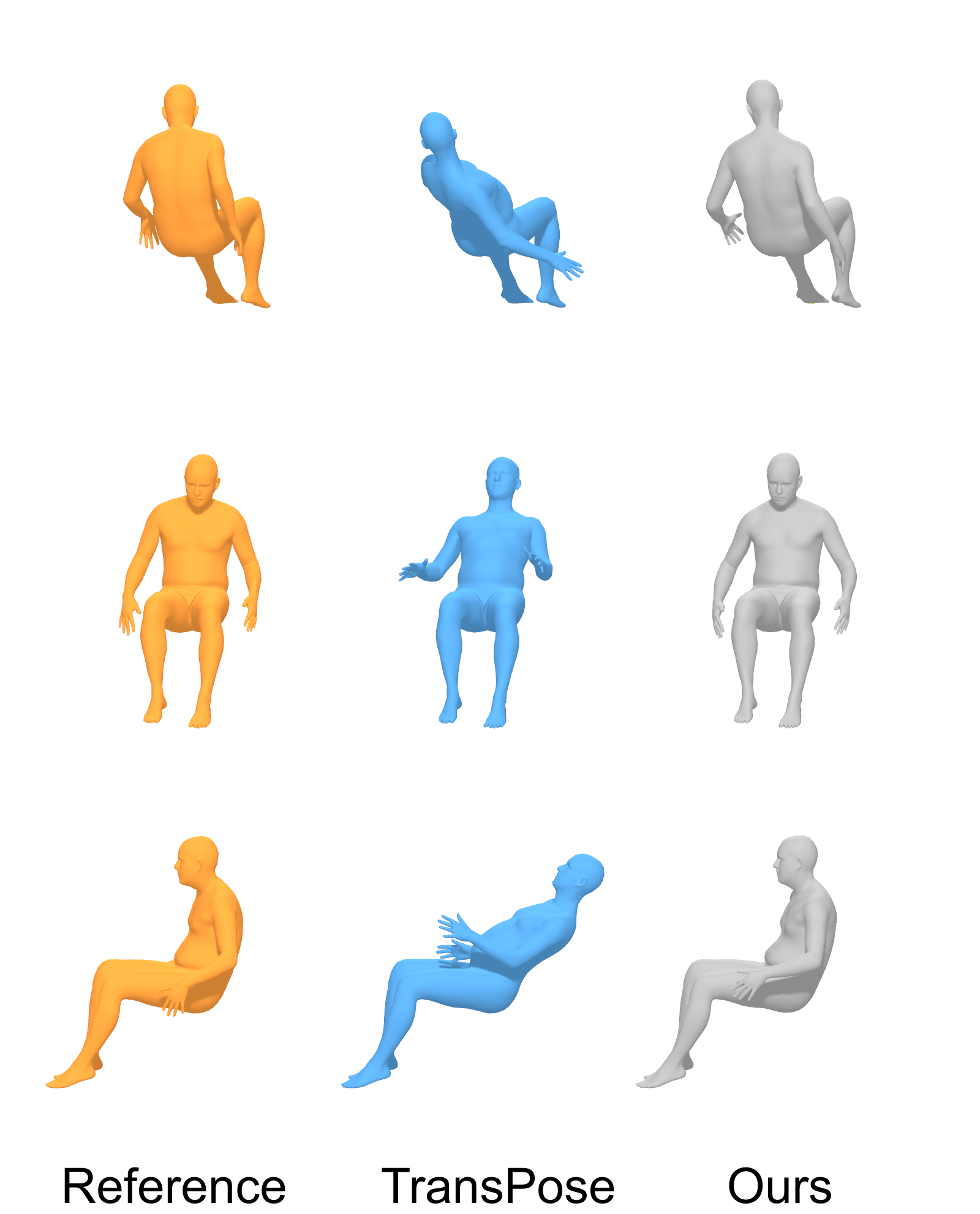}
         \caption{Wheelchair locomotion.}
         \label{fig:qual_chair}
     \end{subfigure}
        \caption{Qualitative comparison of pose estimation results between TransPose and our three-stage biRNN model.}
        \Description{This figure illustrates a collection of pose estimation results from TransPose and WheelPoser, along with the ground truth. The left side displays the pose estimation results for arm motions, indicating that WheelPoser captures these motions more accurately than previous systems. The middle section demonstrates results for torso rotations, showing that WheelPoser captures the rotation angle with greater precision. The right side presents pose estimation results for wheelchair locomotions, revealing that WheelPoser accurately captures both the pelvis joint movements and the wheelchair propulsion and rotation motions.}
    \label{fig:qual_comparison}
\end{figure*}

\begin{table*}[h!]
\renewcommand{\arraystretch}{1.1}
\centering
\resizebox{\textwidth}{!}{%
\begin{tabular}{ccccccc}
\hline
\rule{0pt}{9pt}
\textbf{Models}         & \textbf{Ang Err (deg)} & \textbf{Pos Err (cm)} & \textbf{Mesh Err (cm)} & \textbf{Jitter $\mathbf{(10^2/m^3)}$} & \textbf{Wrist Pos Err (cm)} & \textbf{Elbow Pos Err (cm)}     \\ \hline
IMUPoser~\cite{mollyn2023imuposer}   & 45.20 (10.07)  & 18.27 (5.29)  & 39.13 (5.71)   & 17.59 (26.81) &  24.27 (10.27) & 15.59 (6.69)  \\
TransPose~\cite{yi2021transpose}  & 44.31 (5.97)  & 19.05 (4.77)  & 40.16 (3.92)   & 3.34 (5.72) &23.20 (9.21) & 17.15 (6.79)      \\
PIP~\cite{yi2022physical}   & 63.97 (20.07)  & 23.09 (6.11)  & 49.44 (7.80)   & 3.13 (23.03) 
& 28.42 (12.56) & 22.76 (9.23)\\
TIP~\cite{jiang2022transformer}        & 44.19 (6.12) & 18.56 (3.87) & 40.34 (3.52)             &8.45 (8.68) &21.34 (7.42) & 14.91 (5.87)           \\
Single-stage biRNN (ours)  & 15.09 (5.55)  & 7.12 (3.15)  & 8.89 (2.92)   & \textbf{3.09 (4.65)} & 13.56 (6.19) & 11.38 (5.53)\\
Three-stage biRNN (ours) & \textbf{14.23 (5.26)}  & \textbf{6.71 (2.95)}  & 8.09 (2.72)   & 4.13 (6.04) & \textbf{12.76 (5.81)} & \textbf{11.15 (5.16)}\\
Transformer-based (ours)       & 14.46 (5.48)  & 6.96 (3.30)  & \textbf{7.88 (2.74)}   & 19.39 (17.72) & 12.81 (6.58) & 11.66 (6.10) \\ \hline
\end{tabular}
}
\vspace*{2mm}
\caption{Results for upper body pose estimation across various models: mean (std). We compare our candidate models with prior pose estimation models that utilize sparse IMUs, including IMUPoser~\cite{mollyn2023imuposer}, TransPose~\cite{yi2021transpose}, PIP~\cite{yi2022physical}, and TIP~\cite{jiang2022transformer}.}
\label{tab:kinematics_results_online}
\end{table*}

\subsection{Evaluation of Kinematics Module}\label{kinematics}
The quantitative comparisons of the different kinematics modules are shown in Table~\ref{tab:kinematics_results_online}, where the mean and standard deviation for each metric are presented. 
In addition to assessing the performance of the three proposed kinematics modules, we also evaluate four state-of-the-art (SOTA) models designed for people without motor impairments, including IMUPoser~\cite{mollyn2023imuposer}, TransPose~\cite{yi2021transpose}, PIP~\cite{yi2022physical}, and TIP~\cite{jiang2022transformer}. We use the officially released versions provided by the authors and test them on data from the two full-time wheelchair users. Given that most existing models necessitate a sensor setup of 6 IMUs (with two additional sensors on the legs or thighs), we compensate for missing IMU data from the legs using the IMU data of the wheelchair and solely compare the upper body pose estimation results to ensure fair comparisons.

\subsubsection{SOTA Models versus Our Candidate Models}
First, it is evident that all existing models trained and fine-tuned solely on datasets from individuals without motor impairments are incapable of effectively predicting on-wheelchair motions. This demonstrates a substantial disparity between on-wheelchair motion patterns and those present in existing datasets collected in ambulatory-dominated scenarios, such as standing and walking.
In contrast, our proposed models consistently exhibit significantly better performance than existing models, demonstrating the effectiveness of our proposed data synthesis method, pose estimation pipeline, and the WheelPoser-IMU dataset.

Additionally, we visually compare the pose estimation results between our proposed kinematics modules and SOTA models to provide an intuitive understanding of the sources of error and model differences. For presentation clarity, we showcase the results of our three-stage biRNN-based kinematics module alongside TransPose. Additionally, we present the reference and predicted poses in a full-body manner by matching the lower body joint angles to the ground truth, thereby providing an intuitive demonstration of the pelvis joint angle. 

A major source of difference arises from the estimation of pelvis orientation. Specifically, as illustrated in Figure~\ref{fig:qual_comparison}, models trained and fine-tuned solely on ambulatory motion consistently exhibit bias towards pelvis orientation in standing poses. In contrast, our models can compensate for differences in pelvis orientation. Additionally, wheelchair users often need to perform overhead reaching and far-reaching actions in their everyday lives due to their sitting configuration and environmental accessibility barriers. However, as depicted in Figure~\ref{fig:qual_arm}, models trained exclusively on ambulatory motions display biases in shoulder and arm bending angles and fail to predict straight arm motions accurately. In contrast, our models demonstrate better performance in capturing this essential aspect of wheelchair users' motion. Similarly, wheelchair users often need to rotate or bend their torso for essential activities like pressure relief. In response, our models exhibit improved performance in predicting corresponding trunk joint angles, as shown in Figure~\ref{fig:qual_body}.

Another significant difference between our model and existing models for individuals without motor impairments lies in pose prediction during wheelchair locomotion, such as propulsion and turning. As indicated in Figure~\ref{fig:qual_chair}, our models exhibit significantly better performance in predicting the pose of upper extremities and torso during such motions. More comparisons can be found in our Video Figure.

\begin{table*}[t]
\centering
\renewcommand{\arraystretch}{1.1}
\resizebox{0.95\textwidth}{!}{%
\begin{tabular}{ccccccc}
\hline
\rule{0pt}{9pt}
\textbf{Models}        & \textbf{Ang Err (deg)} & \textbf{Pos Err (cm)} & \textbf{Mesh Err (cm)} & \textbf{Jitter $\mathbf{(10^2/m^3)}$} & \textbf{Wrist Pos Err (cm)} & \textbf{Elbow Pos Err (cm)}     \\ \hline
TransPose~\cite{yi2021transpose}  & 44.31 (5.97)  & 19.05 (4.77)  & 40.16 (3.92)   & 3.34 (5.72) &23.20 (9.21) & 17.15 (6.79)      \\
PIP~\cite{yi2022physical}   & 63.97 (20.07)  & 23.09 (6.11)  & 49.44 (7.80)   & 3.03 (23.03) 
& 28.42 (12.56) & 22.76 (9.23)\\
Kinematics Only & 14.23 (5.26)  & 6.71 (2.95)  & 8.09 (2.72)   & 4.13 (6.04) & 12.76 (5.81) & 11.15 (5.16)\\
With physics & 14.30 (5.31)  & 6.74 (2.98)  & 8.10 (2.73)   & \textbf{2.71 (4.21)} & 12.86 (5.90) & 11.19 (5.21)\\
\hline
\end{tabular}
}
\vspace*{2mm}
\caption{Results of the physics-based optimization module. Here we compare the effectiveness of our module against PIP which focused on foot-ground contact modeling. WheelPoser's physics module reduces jitter error by more than 30\%.}
\label{tab:physics_results_online}
\end{table*}


\subsubsection{Results Across Our Candidate Models}
On comparing the performance of the three variations of our proposed kinematics module, the single-stage biRNN-based model shows the lowest jitter error, the transformer-based model exhibits the lowest vertex error, while the three-stage biRNN-based model outperforms the others across the remaining metrics. The enhanced performance of the three-stage biRNN in estimating joint angles and positions aligns with prior research on people without motor impairments~\cite{yi2021transpose}, highlighting the efficacy of utilizing joint positions as an intermediate representation for learning complex human motion priors in on-wheelchair movements. We also observe a significantly high jitter error from the transformer-based model, similar to TIP~\cite{jiang2022transformer}. Furthermore, the three-stage biRNN-based model presents a higher jitter error compared to the single-stage model. This difference may arise from the presence of noisy joint position outputs in the early-stage networks of our model, as opposed to the end-to-end mapping from IMU measurements to pose employed in the single-stage biRNN model.
Overall, given the enhanced pose estimation performance and relatively low jitter error of the three-stage biRNN-based kinematics model, we select it as the kinematics model for WheelPoser and further evaluation of the physics-based optimization module.

\subsection{Evaluation of Physics Module}\label{physics}
We assess the effectiveness of our physics-based optimization module in reducing the jitter error of the predicted pose generated by the kinematics module. The detailed results are shown in Table~\ref{tab:physics_results_online}.
Here, we revisit the performance metrics of TransPose and PIP for comparison, as TransPose is a kinematics-only model, whereas PIP incorporates physics-based modeling for foot-ground contact in ambulatory motion.
Notably, PIP achieves much worse pose estimation results for on-wheelchair motions compared to its kinematics-only counterpart. We attribute this to its focus on modeling foot-ground contact, which is not directly applicable to and actually hinders pose estimation performance in wheelchair-use scenarios.
Specifically, as PIP predicts foot-ground contact probability using a model trained on ambulatory motions, it is inherently biased towards predicting one or both feet in contact with the ground. These probabilities are then used to refine the pose with a physics-based optimizer, where errors in contact probability lead to errors in pose. However, during wheelchair locomotions, the feet ``slide'' with respect to the ground, which PIP fails to predict accurately, due to this bias. Additionally, PIP optimizes the pose in the global frame, where the ground and footrest should be considered as different objects, leading to further errors. Taken together, these factors cause their approach to fail in wheelchair usage cases. 
In contrast, our physics-based optimization module is designed specifically for on-wheelchair motions and significantly reduces the jitter error by more than 30\% while maintaining performance across other metrics.

\begin{figure}[h!]
    \centering
    \includegraphics[width=0.95\linewidth]{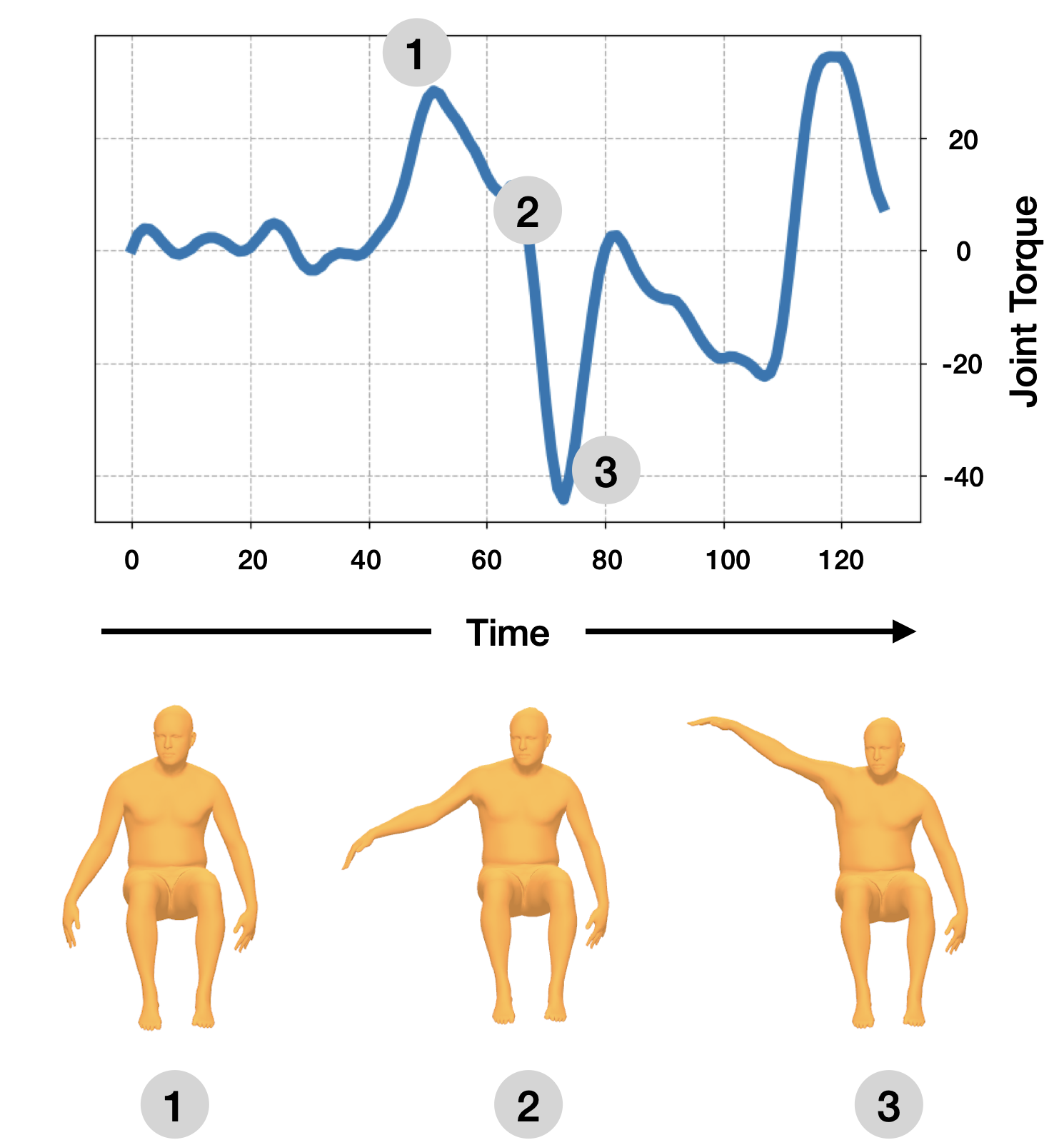}
    \caption{Visualization of estimated joint torques of the right shoulder joint and its corresponding motion sequences.}
    \Description{This figure displays the estimated joint torque information during an upper body motion. The left side shows a motion sequence of raising the left arm. On the right side, a line chart presents the estimated joint torque. The joint torque is high at the beginning of the motion due to acceleration, drops to around zero in the middle when acceleration decreases, and increases to negative values at the end to decelerate the arm movement.}
    \label{fig:joint_torque}
\end{figure}

The inclusion of a physics module offers other advantages. For instance, we can now estimate the joint torques of the user, which are critical to applications like upper extremity health monitoring, which we will discuss in Sec.~\ref{collaborative_care}. Figure~\ref{fig:joint_torque} shows exemplary estimated joint torques. 
We can see that in order for the user to accelerate their arm for the lifting motion, their right shoulder produces a large positive torque (1) and as the arm reaches its highest point, the right shoulder slows the arm down by producing a large negative torque (3). And the estimated joint torque is lower when the user's arm is moving at a steady speed (2).

\subsection{Performance across Motion Types}\label{types}
\begin{figure*}[b]
    \centering
    \includegraphics[width=\textwidth]{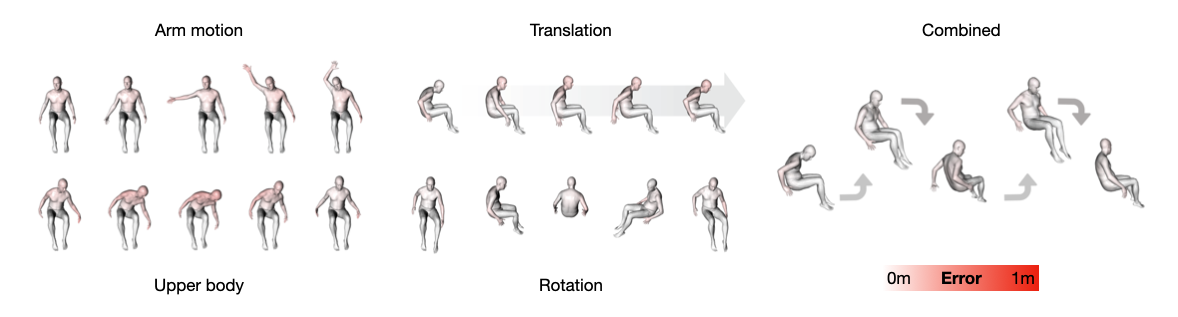}
    \vspace{-8pt}
    \caption{Sample estimated poses of our model categorized by its types. The rendered pose is colored based on the mesh error.}
    \Description{This figure showcases sample pose estimation results from WheelPoser across different categories of motions. It includes five types of motions: arm motion, upper body motion, translation, rotation, and combined motions. The arm motion segment illustrates a sequence where a person raises their right arm. The upper body segment depicts a person leaning to their left. The translation segment demonstrates a sequence of pushing forward. The rotation segment shows a person turning in place on a wheelchair. Finally, the combined motion segment captures a person performing a turn while moving forward.}
    \label{fig:motion_type}
\end{figure*}

In this section, we delve into a detailed analysis of our model's performance across various wheelchair motion types. Specifically, we categorize our collected wheelchair motion types into five distinct groups, comprising: 1) arm motion and 2) upper body motion as detailed in Sec.~\ref{collection_protocol}, 3) translational wheelchair locomotion (pushing forward and backward), 4) rotational wheelchair locomotion (turning in place and executing a pivot turn), and 5) combined wheelchair locomotion (pushing forward in a circle, pushing backward in a circle, and turning while moving forward). The results of these analyses are presented in Table~\ref{tab:motion_type_results_online}.
Specifically, we observe that our model demonstrates comparable performance on wheelchair location motions, including translational, rotational, and combined wheelchair motions, yet experiences a drop in performance on stationary motions such as arm and upper body movements. 
For instance, arm motions demonstrate the largest wrist position error and joint angle error, and upper body motions demonstrate the largest error in joint position and body mesh.
We attribute these increased errors in arm motions to biases in the elbow and shoulder bending angles present in AMASS dataset. 
Specifically, straight arm movements and overhead arm motions are underrepresented in AMASS. Yet, they constitute a significantly larger portion of our dataset, reflecting typical wheelchair use characteristics. 
Regarding the higher errors found in upper body motions, we believe they are tied to the low accelerations of the user's head during these movements. Specifically, as our model largely relies on the single IMU positioned on the user's head to infer the trunk motion, the low accelerations hinder the model's ability to accurately predict joint positions and consequently the body mesh.
For qualitative understanding, we also showcase samples of estimated pose colored based on mesh error in Figure~\ref{fig:motion_type}.
\begin{table*}[t]
\centering
\resizebox{0.95\textwidth}{!}{%
\begin{tabular}{ccccccc}
\hline
\rule{0pt}{10pt}
\textbf{Motion Types}         & \textbf{Ang Err (deg)} & \textbf{Pos Err (cm)} & \textbf{Mesh Err (cm)} & \textbf{Jitter $\mathbf{(10^2/m^3)}$} & \textbf{Wrist Pos Err (cm)} & \textbf{Elbow Pos Err (cm)}     \\ \hline
Arm motion  & 16.60 (5.41)  & 7.61 (3.36)  & 9.70 (2.52)   & 1.65 (2.61) &14.81 (6.88) & 11.71 (5.66)      \\
Upper body   & 15.73 (5.74)  & 7.68 (3.12)  & 10.44 (3.10)   & 1.43 (1.76) 
& 13.56 (5.73) & 11.11 (5.30)\\
Translation & 13.57 (4.21)  & 5.98 (2.77)  & 8.18 (3.16)   & 2.83 (3.55) & 10.83 (5.23) & 9.91 (4.64)\\
Rotation & 13.51 (4.87)  & 6.61 (2.64)  & 7.56 (2.20)   & 2.81 (3.91) & 12.97 (5.45) & 11.74 (4.81)\\
Combined & 14.30 (5.31)  & 6.74 (2.98)  & 8.10 (2.73)   & 3.21 (4.33) & 11.91 (5.60) & 10.56 (4.90)\\
\hline
\end{tabular}
}
\vspace*{2mm}
\caption{Results for upper body pose estimation across different motion types.}
\label{tab:motion_type_results_online}
\end{table*}

\subsection{Live Demo}
We further implement a real-time live pose estimation system using 4 Movella DOTs and Unity. Our system runs in real-time at 60~fps on a Dell G16 Laptop with an Intel Core i7 12700H CPU and an NVIDIA 3060 mobile GPU. From the live demos, we can see that our system achieves high accuracy and real-time pose estimation for various motion types (refer to Video Figure). It is able to effectively handle diverse lighting conditions, visual occlusions, as well as different environmental contexts. It also demonstrates promising potential to generalize to different wheelchair models. We note that our frame rate is bottlenecked by the frame rate (60~fps) of Movella DOTs. Processing a window of 26 IMU frames in real-time (details in Sec.~\ref{Eval}) takes 10.57~ms on our hardware, meaning WheelPoser could potentially run live at over 90~fps with faster IMUs.

\section{Example Applications}\label{applications}
We envision WheelPoser to enable a wide range of applications across various fields by making IMU-based pose estimation inclusive and practical for everyday uses. The following application concepts highlight how WheelPoser can be used by both full-time and novice wheelchair users as well as other relevant stakeholders in the future.

\subsection{Facilitating Wheelchair Skill Training}
The World Health Organization (WHO) Guidelines on the Provision of Wheelchairs~\cite{world2023wheelchair} necessitates the importance of receiving wheelchair skill training for wheelchair users. However, the manner in which people receive their wheelchairs varies widely and many wheelchair users report having limited wheelchair skill training and education resources around~\cite{li2023breaking, kirby2016wheelchair}. Applications that aim to facilitate wheelchair skill training in both clinical and community contexts can be built on top of our pose estimation system.
For instance, the propulsion patterns used by wheelchair users when self-propelling are critical parameters that can describe the quality of their wheelchair propulsion techniques and significantly influence the likelihood of experiencing an upper extremity injury~\cite{paralyzed2005preservation}. Despite the limited skill training resources, clinic-based training further suffers from the cognizant effect, and existing systems can only classify basic wheelchair propulsion patterns, such as semi-circular and half-moon patterns~\cite{chen2018toward, french2008classifying, herrera2018towards}. Our practical pose estimation system could track more fine-grained propulsion trajectories in wheelchair users' everyday lives, allowing for more personalized training guidance to facilitate the adoption of efficient propulsion techniques. Similarly, wheelchair users' poses during wheelchair transfers can be tracked and analyzed for safe transfer technique training. Further, alerts based on best practices for recognized pressure relief frequency and timing can be offered.

\subsection{Inclusive Personal Informatics}
Personal Informatics (PI) systems aim to provide users with actionable, data-driven insights from personally relevant information, thereby enhancing their quality of life~\cite{kersten2017personal}.
Although these systems have been extensively researched across diverse populations and contexts, individuals with motor impairments have received significantly less attention, leaving many of their crucial needs unmet~\cite{motahar2022review}. Specifically, for wheelchair users, previous research indicates that they can greatly benefit from monitoring specific activities like pressure relief and wheelchair usage as well as general upper body motions~\cite{li2023breaking}. In this case, tracking a wheelchair user’s pose could serve as an intrinsic indicator of human activity and be particularly informative for tracking a wide range of activities, in contrast to previous work on single-purpose systems~\cite{barbareschi2020understanding, ma2017activity}. For instance, by analyzing the pose of a wheelchair user, we can detect the occurrence and type of pressure relief actions, which are critical for preventing pressure ulcers~\cite{motahar2022identifying}. Additionally, monitoring the elbow's position over time can provide valuable data for assessing a wheelchair user's range of motion and supporting wheelchair users' upper extremity health management~\cite{li2023breaking}.

\subsection{Novel Interaction and Gaming Experiences}
Practical pose tracking unlocks exciting new possibilities for novel interaction and gaming experiences for wheelchair users. For instance, by enabling accurate pose estimation without using cameras, WheelPoser addresses privacy concerns while facilitating the implementation of diverse input methods, such as on-body, in-air, and on-wheelchair gestures as explored by Bilius et al.~\cite{bilius2023understanding}. Additionally, it broadens the design space for developing new, intuitive body-based interaction techniques, allowing for the creation of novel interaction and locomotion techniques in VR/AR environments using natural poses or physical wheelchair locomotion, reducing the need for inaccessible controllers~\cite{mott2020just}.
Similarly, a wheelchair user's pose provides rich contextual information that could drive implicit interactions in applications like smart home systems or autonomous wheelchair navigation~\cite{jang2022should}.
Furthermore, WheelPoser's real-time and on-the-go pose estimation capabilities open up new avenues for inclusive gaming experiences, expanding the reach of both rehabilitation-focused and recreational games to more diverse settings, such as outdoor environments and mobile contexts.

\subsection{Enhanced Data-Driven Collaborative Care} \label{collaborative_care}
Expanding beyond the aforementioned application domains, we envision WheelPoser enabling a variety of applications involving other stakeholders, such as rehabilitation scientists, physical therapists, assistive technology specialists, occupational therapists, and caregivers. For instance, the pose and joint torque of wheelchair users outside of clinics hold immense potential to advance rehabilitation scientists' understanding of the pathology of issues like upper extremity injuries~\cite{mercer2006shoulder}. It can also inform physical therapists about the effectiveness of dosed treatment and patient compliance~\cite{li2023breaking}. Furthermore, metrics like the count of overhead reaching can reflect the accessibility of certain physical environments and inform occupational therapists about environmental modifications. Additionally, the analysis of a user's pose while in a wheelchair extends its utility to assistive technology specialists on issues like wheelchair fitting. Specifically, user pose data may pinpoint issues related to wheelchair ergonomics and therefore provide a foundation for tailoring wheelchair designs to individual user needs.

\section{Discussion and Future Directions}
Like any system, WheelPoser has limitations. Here, we reflect on the evaluation results of WheelPoser, discuss typical failure cases, and explore future opportunities for advancing pose estimation for wheelchair users. 

\subsection{Generalization}
As a data-driven approach, learning-based human pose estimation faces challenges with generalization. In our work, we have shown that our proposed data synthesis and pose estimation pipeline demonstrates strong generalizability as it enables significant improvements in pose estimation across different model architectures compared to SOTA methods (see Table~\ref{tab:kinematics_results_online}). Additionally, our trained model is able to generalize well to unseen data as it achieves good quantitative and qualitative results on the held-out test split containing full-time wheelchair users' data. 
\begin{table*}[h]
\centering
\renewcommand{\arraystretch}{1.2}
\resizebox{\textwidth}{!}{%
\begin{tabular}{ccccccc}
\hline
\textbf{Models}         & \textbf{Ang Err (deg)} & \textbf{Pos Err (cm)} & \textbf{Mesh Err (cm)} & \textbf{Jitter $\mathbf{(10^2/m^3)}$} & \textbf{Wrist Pos Err (cm)} & \textbf{Elbow Pos Err (cm)}     \\ \hline
Leave-one-subject-out & 14.30 (5.31)  & 6.74 (2.98)  & 8.10 (2.73)   & 2.71 (4.21) & 12.86 (5.90) & 11.19 (5.21)\\
Without Full-time Wheelchair Users & 16.15 (5.53)  & 7.65 (3.11)  & 9.57 (3.05)   & 2.82 (4.77) & 14.89 (6.11) & 12.65 (5.36)\\
\hline
\end{tabular}
}
\vspace{2pt}
\caption{Comparison of model performance when fine-tuned with data from full-time wheelchair users versus exclusively with data from participants without motor impairments.}
\label{tab:am_only_results}
\end{table*}

\begin{figure}[b]
 \centering
\includegraphics[width=0.8\linewidth]{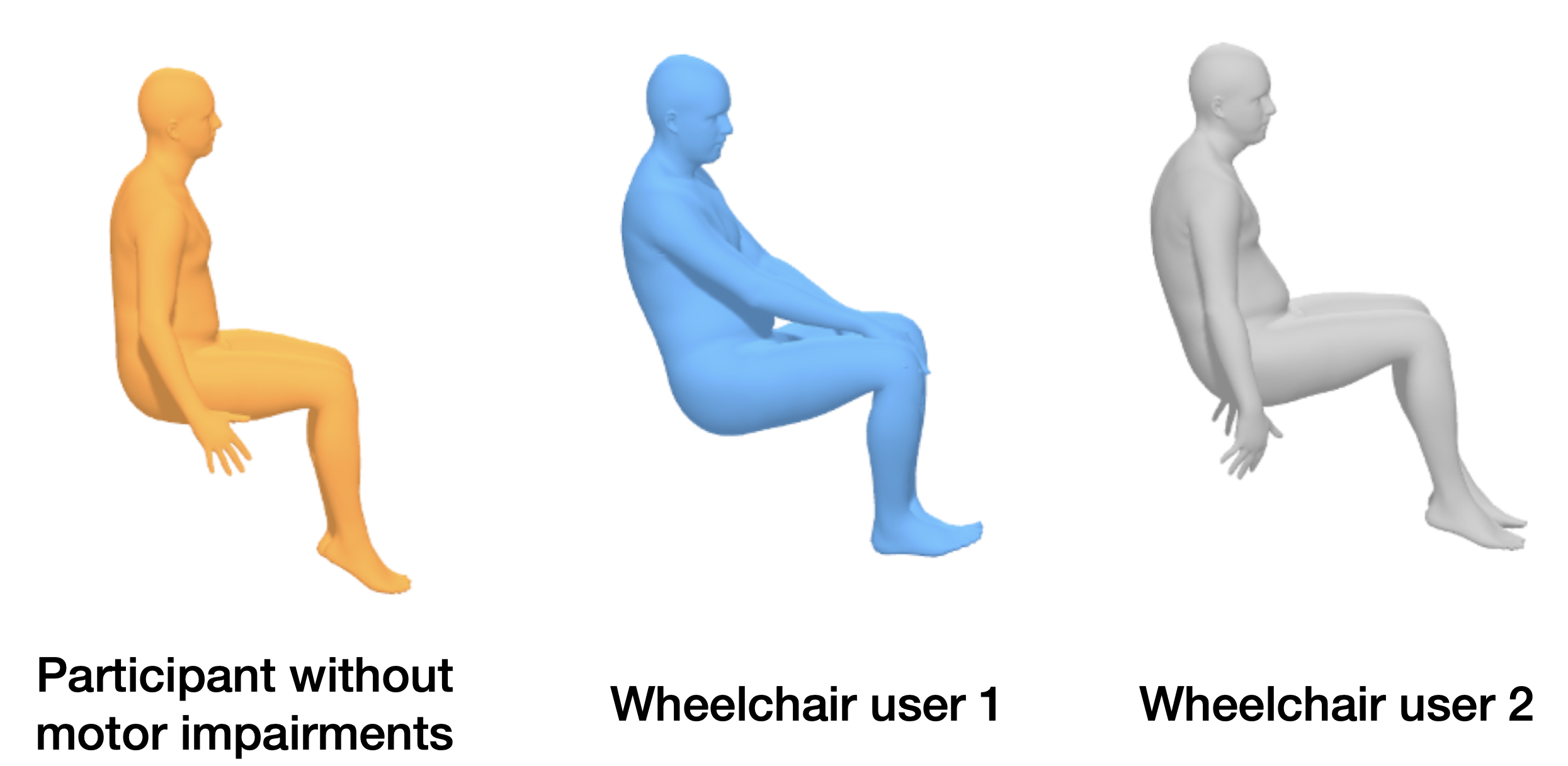}
\caption{Sample sitting postures of our participants.}
\Description{This figure demonstrates the sitting postures of our participants. The image on the left is of individuals without motor impairments, and the two images on the right are of full-time wheelchair users. The figure illustrates that the sitting posture of people without motor impairments slightly differs from those of full-time wheelchair users.}
\label{fig:sitting pose}
\end{figure}

We further assess the generalizability of our dataset by fine-tuning a three-stage biRNN based model exclusively on data from participants without motor impairments and evaluating its performance on the data from full-time wheelchair users. In Table~\ref{tab:am_only_results}, we observe a slight performance drop, with error metrics increasing by about 11\% on average. Still, this data significantly improves pose estimation performance compared to existing SOTA models trained and fine-tuned solely on ambulatory motion data. This result is promising given the prevalence of simulated data to overcome difficulties in recruiting disabled users to help narrow performance gaps. However, we acknowledge and emphasize the importance of collecting data from \textit{real} members of intended user groups.

Careful examination of the ground truth poses reveals a possible gap between our participants without motor impairments and full-time wheelchair users. Specifically, many wheelchair users lack trunk control due to their physical impairments, and their default sitting posture may exhibit different characteristics than those of people without motor impairments, as shown in Figure~\ref{fig:sitting pose}. This difference may have led to increased error in pose estimation, especially in body mesh error. 
Additionally, based on our observation, full-time wheelchair users generally demonstrated smoother wheelchair locomotion (e.g. higher stroke efficiency), likely due to extensive usage experience.
More broadly, we acknowledge that each wheelchair user is characterized by their own body shape, posture, flexibility, mobility differences, and wheelchair types, all of which are likely to influence their motion patterns and subsequently, the IMU signals. To tackle this problem, we attempted to synthesize pose and motion data from relevant online videos (e.g., wheelchair skill training videos) using state-of-the-art (SOTA) models for human pose estimation from monocular videos—4D Human~\cite{goel2023humans}. However, we obtained poor performance in pose reconstruction wheelchair users, and the generated data exhibited poor accuracy and was temporally inconsistent, ultimately proving unhelpful in improving the performance of our model.

\subsection{Failure Cases}

\begin{figure}[b]
 \centering
\includegraphics[width=0.8\linewidth]{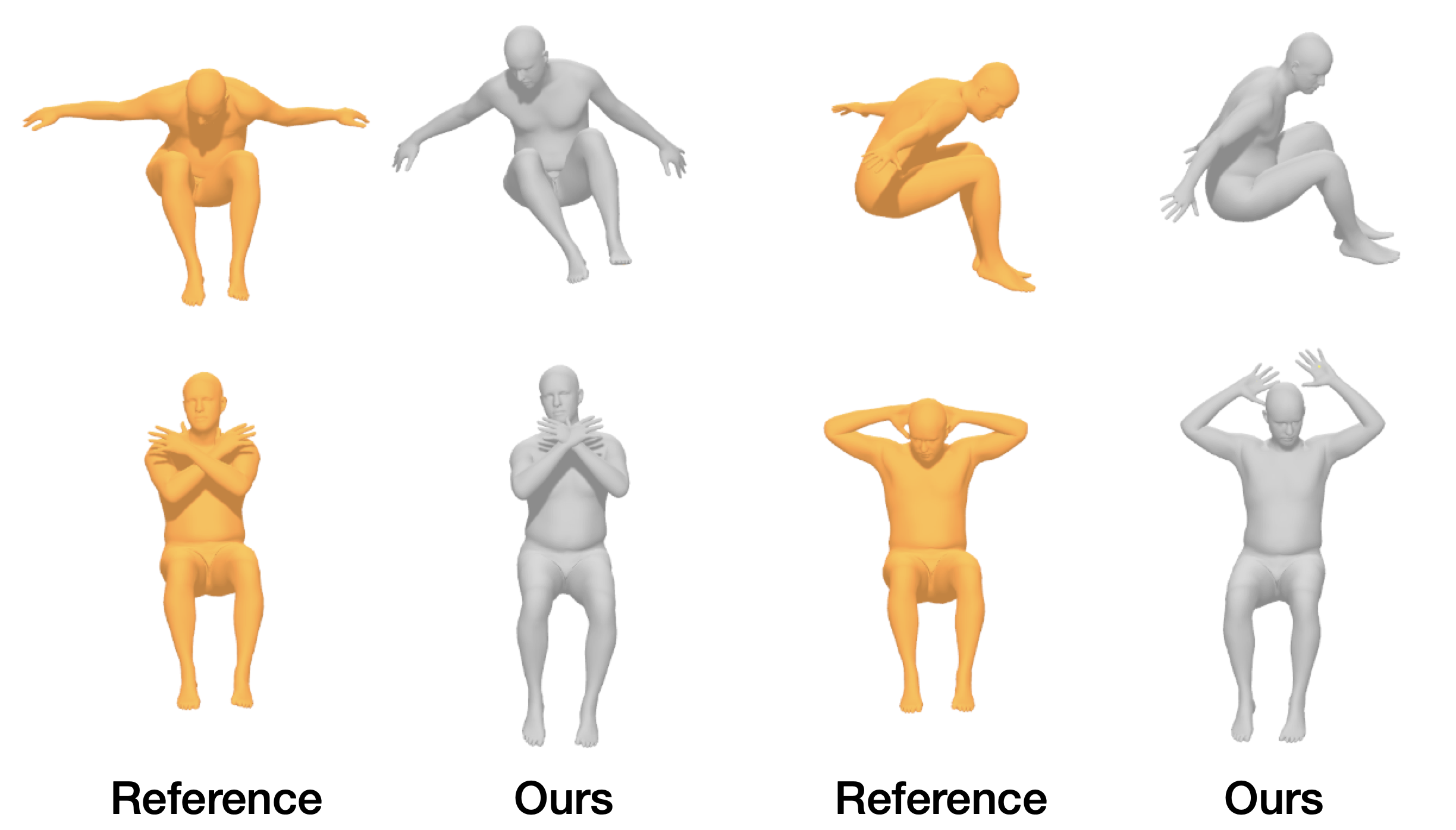}
     \caption{Typical failure cases of our model.}
     \Description{This figure highlights certain pose estimation results from our system that are underperforming. The top row displays a leaning forward motion, while the bottom row depicts arm crossing motions. The predicted poses show less accurate torso and elbow angles compared to the ground truth.}
    \label{fig:failure_cases}
\end{figure}

Our system relies on IMUs on the user's head and forearms to predict the user's pose when their wheelchair remains stationary. Additionally, our model utilizes IMU acceleration to address the motion ambiguity introduced by our sparse-IMU sensor setup. Taken together, these factors result in challenging cases when the user slowly moves their upper body while both of their hands remain stable, such as the leaning forward motion illustrated in Figure~\ref{fig:failure_cases} (top row). Similarly, precise positioning of the user's arm can be difficult as it heavily relies on the acceleration data and our model does not explicitly account for different bone lengths among users, influencing the acceleration profile. Consequently, our model underperforms in motions that involve close coordination of both arms and may result in mesh penetration, as shown in Figure~\ref{fig:failure_cases} (bottom row).

More broadly, an inertial sensor relies on its magnetometer to facilitate orientation measurement, which can be affected by the magnetic field of the environment. As we are attaching one IMU to the axle of the wheelchair, the material used to manufacture the wheelchair can potentially affect the IMU measurement and, therefore, lead to errors. Additional errors may also be introduced if the user has a relative position change from their wheelchair, as we have no direct way of measuring the user's pelvis position.

\subsection{Facilitating Marker-based Motion Capture}
In our study, we employed a commercial marker-based motion capture system to collect human motion ground truth. This approach required participants to visit our motion capture studio in person, which imposed logistical challenges to both the participants and the research team. Given the limited accuracy of existing RGB camera-based solutions for tracking wheelchair users' poses, commercial motion capture systems are likely to remain necessary for motion ground truth collection in future research. Here, we share several lessons learned from our experience to inform similar endeavors.
First, motion capture with marker-based systems involves an extensive setup. A traditional full-body maker setup, for example, requires at least six markers on the user's back. In this work, we attached part of the markers to the wheelchair backrest to avoid visual occlusion based on each participant's wheelchair setup. Here, we found it beneficial to request a few photos of participants in their wheelchairs prior to data collection and plan the potential marker placements, which significantly reduced the length of the setup process. Additionally, different considerations and adjustments may be needed for power wheelchair users, as power wheelchairs tend to provide more trunk and head support, which can introduce even more visual occlusion.
Furthermore, coordinating with participants beforehand to review the motions to be collected is critical. We found this preparatory step to be essential in ensuring participant comfort and facilitating the planning of necessary adjustments. In our study, we were able to adjust the planned motions for W14 prior to data collection based on his trunk control capacity. In addition, both the marker-based motion capture systems and our system require calibration motions. In this study, we opted for a T-pose calibration, which our participants were comfortable with and able to perform. Future studies might explore more accessible calibration procedures that consider factors such as the wheelchair users' range of motion and the type of wheelchair, among others.

\subsection{Future Opportunities}
\subsubsection{Towards Even More Practical and Unobtrusive Motion Capture}
In this work, we demonstrated the feasibility of using four IMUs for accurate upper body pose estimation for wheelchair users. Compared to commercial systems that require dense-IMU setups, WheelPoser represents a crucial step toward inclusive, practical, and unobtrusive motion capture for wheelchair users. Specifically, our sensor setup effectively lowers accessibility barriers and our pose estimation exhibits significant improvement over SOTA methods. Further, because of the improved practicality and accessibility, WheelPoser enables a wide range of applications spanning diverse fields and multiple stakeholders, as discussed in Sec.~\ref{applications}.
Despite being much more practical and unobtrusive than previous systems, we acknowledge that wearing and managing four IMUs can still be challenging for some users. 
Future work may explore ways to further reduce the number of required IMUs or achieve flexible sensor setups that afford user customization. Additionally, our sensor setup can be implemented using IMUs from devices that people already carry daily, such as smartwatches, earbuds, and smartphones. Future work may explore developing pose estimation systems for wheelchair users utilizing these devices to achieve more unobtrusive motion capture, similar to previous work with individuals without motor impairments~\cite{mollyn2023imuposer}.
Furthermore, a wheelchair offers ample opportunities for instrumentation, and previous studies have demonstrated that many wheelchair users prefer devices in the chairable form factor~\cite{carrington2014wearables}. Future research may explore ways to improve the pose estimation accuracy by incorporating sensor fusion techniques and integrating additional sensor types, such as ultra-wideband (UWB) anchors~\cite{devrio2023smartposer, li2022travelogue} and pressure sensors~\cite{luo2021intelligent}, directly into the wheelchair.
\subsubsection{Advanced Physics-based Modeling for Pose Estimation and Wheelchair Kinematics Tracking}
Another important direction for future research involves designing and integrating more sophisticated physics-based modeling into the pose estimation pipeline. In our current work, we have shown that physics-based modeling of general human motion and seating contact significantly enhances the smoothness of predicted on-wheelchair motions. Future studies could investigate tracking and modeling the complex hand-pushrim contact, which could lead to more physically plausible pose estimations. Additionally, modeling the interaction between the wheelchair and the ground may improve the accuracy of wheelchair kinematics tracking, offering advancements over current kinematics-only methods. 

\subsubsection{Towards Large-Scale Inclusive Motion Capture Datasets}
As learning-based approaches become more prevalent across problem domains, the creation of inclusive datasets becomes essential to mitigate bias in machine learning models~\cite{guo2020toward, olugbade2022human}. Currently, most open-source motion capture datasets are collected predominantly from individuals without motor impairments, featuring typical ambulatory motions such as standing and walking. In this work, we take an initial step towards developing large-scale wheelchair motion capture datasets by contributing a dataset that includes 167 minutes of on-wheelchair motions from 14 participants. 
Future work should encompass a broader spectrum of wheelchair users, including not only those who use manual wheelchairs but also power wheelchair users, as well as both novice and part-time users alongside full-time users.
We recognize that building such datasets involves significant labor and financial challenges that extend beyond the capacity of individual labs or organizations. Therefore, we advocate for sustained and large-scale collaborations and call for active participation from the wider research community to join us in these efforts.

\subsubsection{From Wheelchairs to Other Mobility Aids} 
More broadly, tracking and understanding users' motion in relation to their mobility aids offers valuable insights and opens up a wide range of research possibilities. In this study, we demonstrated the feasibility of tracking wheelchair users' poses using a setup specifically tailored to the characteristics of wheelchair use. Looking ahead, future research could extend to a wide range of mobility aids, such as walking canes and rollators. Each of these aids features distinct dynamics and user motion patterns, presenting unique challenges for pose estimation and data collection. Beyond the technical challenges, however, the applications of this research extend substantial benefits across both digital and physical spaces. For instance, inclusive pose estimation enables inclusive representations in digital spaces (e.g. inclusive avatars~\cite{mack2023towards,zhang2023diary}) that reflect a wider spectrum of user identities and experiences. In addition, detailed analysis of user behavior derived from pose estimation data can facilitate identifying and subsequently addressing accessibility barriers within physical environments. 

\section{Open Source}
To enable others to build upon our system, we make our dataset, architecture, and trained model freely available to the research community at
\url{https://github.com/axle-lab/WheelPoser}.

\section{Conclusion}
We present WheelPoser, a sparse-IMU based wheelchair user tracking system that estimates upper body pose in real-time using only four strategically placed IMUs on the user’s body and their wheelchair. We contribute a pose estimation pipeline tailored to wheelchair users while allowing us to leverage existing high-quality motion capture datasets collected from people without motor impairments. We also contribute the first-of-its-kind WheelPoser-IMU dataset, containing 167 minutes of on-wheelchair motions performed by 2 full-time wheelchair users and 12 participants without motor impairments. Our evaluations demonstrate the performance of WheelPoser in accurately predicting wheelchair user motions, significantly surpassing prior sparse-sensor based approaches. Overall, we believe WheelPoser holds immense potential to unlock a wide range of new application possibilities for wheelchair users, such as inclusive self-tracking experiences, wheelchair skill training support, and enhanced collaborative healthcare, all while using an accessible sensor setup.

\begin{acks}
The authors would like to thank all participants for their time and effort in our study. We extend our gratitude to Justin Macey and Jessica Hodgins for their assistance in collecting the motion capture data. We also appreciate Giorgio Becherini and the Perceiving Systems group at the Max Planck Institute for Intelligent Systems for their help with processing the marker data. Finally, we would like to thank all members of the AXLE Lab at CMU, Chris Harrison, Soyong Shin, and the anonymous reviewers for their valuable suggestions and feedback.
\end{acks}

\bibliographystyle{ACM-Reference-Format}
\bibliography{sample-base}

\appendix

\end{document}